\documentclass[10pt,aps,twocolumn,superscriptaddress,longbibliography, prd]{revtex4-2}

\usepackage{array}
\usepackage{natbib}
\usepackage{amssymb}
\usepackage{amsmath}
\usepackage[utf8]{inputenc}
\usepackage{graphicx}
\usepackage{multirow}
\usepackage{xcolor}
\usepackage{color}
\usepackage{aas_macros}
\usepackage{ulem}
\usepackage{lipsum,afterpage}
\usepackage{hyperref} 
\usepackage{cleveref}
\usepackage{tensor}
\usepackage{mathtools}
\usepackage{physics}
\usepackage{makecell}
\usepackage{cellspace}

\hypersetup{colorlinks=true, linkcolor=blue, citecolor=blue, urlcolor=blue, linktoc=all, linktocpage=true}
	 
\DeclareSymbolFont{eulerfraktur}{U}{euf}{m}{n}
\SetSymbolFont{eulerfraktur}{bold}{U}{euf}{b}{n}
\DeclareSymbolFontAlphabet{\mathfrak}{eulerfraktur}
\DeclareMathSymbol{\varvarfrakh}{\mathord}{eulerfraktur}{'110}

\newcommand{\Chr}[3]{\Gamma{^#1_{#2#3}}}

\newcommand{\DChr}[4]{\Gamma{^#1_{#2#3,#4}}}
\newcommand{\idx}[1]{\indices{#1}}
\newcommand{\idxudd}[3]{\indices{^#1_{#2#3}}}

\newcommand{\Q}[0]{\mathbb{Q}}
\newcommand{\fQ}[0]{f'(\Q)}
\newcommand{\fQQ}[0]{f''(\Q)}

\begin{document}

\title{Neutron stars in $f(\Q)$ gravity}

\author{Lavinia Heisenberg}
\affiliation{Institut für Theoretische Physik, Ruprecht-Karls-Universität Heidelberg, Philosophenweg 16, 69120 Heidelberg, Germany}

\author{Carlos Pastor-Marcos}
\email{pastor\_c@thphys.uni-heidelberg.de}
\affiliation{Institut für Theoretische Physik, Ruprecht-Karls-Universität Heidelberg, Philosophenweg 16, 69120 Heidelberg, Germany}

\date{\today}

\begin{abstract}
We investigate the challenges of constructing neutron star (NS) solutions in $f(\Q)$ gravity, highlighting the importance of treating the affine connection as an active, dynamical component of the theory. 
We begin by clarifying under what conditions standard simplifications —such as the coincident gauge or General Relativity (GR)-like connections— inadvertently lead to GR behavior, even in non-trivial $f(\Q)$ models. Building on previous work in black hole (BH) spacetimes, we adapt the formalism to NS and extend it to non-vacuum configurations. Focusing on two representative models, $f(\Q) = \Q + \alpha \Q^2$ and $f(\Q) = \Q^\beta$, our analysis suggests that, under standard regularity assumptions, solutions with Maclaurin/Laurent-type series recover GR dynamics, pointing to more intricate structures as the likely seat of beyond-GR effects, and reflecting the constraints imposed by the connection’s dynamics on the asymptotic behavior of genuinely beyond-GR solutions. We then formulate the problem as a boundary value problem (BVP) and highlight the numerical pathologies that may arise, together with possible strategies to prevent them. This work aims to provide a concrete framework for future numerical studies and outlines the theoretical consistency conditions required to construct physically meaningful beyond-GR NS solutions in $f(\Q)$ gravity.
\end{abstract}

\maketitle


\section{Introduction}
\label{sec:intro}
\vspace{-0.4em}
Einstein's theory of GR has consistently demonstrated remarkable success in describing gravitational interactions across a vast range of scales, from planetary orbits to the evolution of the Universe itself. However, despite its success, several theoretical and observational challenges have prompted the exploration of alternative gravitational frameworks. Cosmological tensions, BH singularities, the nature of dark matter and dark energy, and the incompatibility of GR with quantum mechanics have all raised questions about the completeness of GR, to the point where these issues have driven interest in reformulating our theory of gravity to explore new gravitational interactions that retain GR's core principles while broadening the spectrum of potential solutions.

In particular, the study of gravitational phenomena through the lens of non-metric geometry is a relatively novel approach that moves beyond the well-established curvature-based interpretation of GR. While traditional formulations of the latter describe gravity as a manifestation of spacetime curvature, structured around the Levi-Civita connection ---which is both metric-compatible and torsion-free---, alternative formulations reveal that gravity can also be framed in terms of torsion or non-metricity \cite{Heisenberg_2019, GeomTrinity, BeltranJimenez:2019odq}. This shift gives rise to the so-called Symmetric Teleparallel Gravity (STG) theories, and in particular, the Symmetric Teleparallel Equivalent of GR (STEGR), where gravity is represented not by curvature but purely by non-metricity, allowing for a framework in which the connection is independent of the metric \cite{Jim_nez_2018, Jim_nez_2018_2, Heisenberg_2019, GeomTrinity}. In STEGR, instead of the Riemann tensor, it is the non-metricity tensor $Q_{\alpha\mu\nu}$ the geometrical object that encodes the gravitational interaction. The non-metricity scalar $\Q$ ---derived from this tensor--- becomes then the gravitational Lagrangian, analogous to the Ricci scalar in GR. In the same vein, Coincident General Relativity (CGR)\cite{Jim_nez_2018}, as the gauge-restricted version of STEGR, maintains flat and torsion-free conditions but allows for a vanishing affine connection, which simplifies the representation of gravitational effects while maintaining consistency with empirical data. 

This non-metric approach is becoming more popular and valuable in modified gravity theories, particularly $f(\Q)$ gravity \cite{heisenberg2023reviewfqgravity}, which extends STEGR by introducing functional forms of the non-metricity scalar $\Q$ in the action \cite{Jim_nez_2018,Heisenberg_2019,  heisenberg2023reviewfqgravity, Heisenberg:2023wgk}. Such non-linear generalization introduces a fundamental change in the sense that the affine connection is no longer a pure gauge quantity that can be entirely removed in the so-called coincident gauge. Instead, it acquires physical significance and becomes a dynamic entity, contributing to the field equations and introducing new degrees of freedom \cite{DAmbrosio:2020nqu,DAmbrosio:2023asf, heisenberg2025countingdegreesfreedommethod}. As a result, additional equations of motion (EoMs) emerge in $f(\Q)$ gravity, giving rise to potentially propagating modes beyond those present in GR. Thus, unlike in STEGR, the connection in $f(\Q)$ gravity cannot be trivially fixed or removed without altering the theory's dynamics, which endows the connection with an active, physically meaningful role in determining gravitational behavior, potentially leading to solutions that differ significantly from those in standard GR \cite{DAmbrosio2021, DAmbrosio2022,  R_nkla_2018,heisenberg2023reviewfqgravity, DAmbrosio:2020nev}.

One of the primary challenges in $f(\Q)$ gravity research lies precisely in understanding how these additional degrees of freedom associated with the connection influence the behavior of gravitational systems. A growing number of studies over the past few years have explored the implications of $f(\Q)$ gravity for static and stationary spherically symmetric configurations, reflecting increasing interest in its potential to describe astrophysical phenomena beyond GR. In many of these works, especially in the early literature, simplifying assumptions on the connection are adopted in order to render the field equations tractable. While such choices have led to a wide range of interesting results, they often suppress the connection's unique contributions to the dynamics, so that the resulting gravitational solutions tend to remain close to their GR counterparts. In practice, such constraints can effectively remove or strongly restrict the new degrees of freedom that are intrinsic to $f(\Q)$ gravity, preventing the theory from fully displaying its potential to describe novel gravitational effects \cite{PhysRevD.106.043509, Jim_nez_2018, DAmbrosio2021, DAmbrosio2022}. 

These studies include many works on BHs such as \cite{Chen_2025,nashed20253dimensionalchargedblackholes,Junior_2023,Junior_2024,Gogoi_2023,Nashed_2024,nashed2025specialndimensionalchargedantidesitter,zhang2025commentblackholesfmathbbq}
, which typically handle trivial connections or assume particularly simple choices, tailored to render the field equations more tractable. Such setups, while often useful and valuable as benchmarks, naturally explore only a subset of the theory's possible dynamical behaviors. More relevant for our purposes, a large body of work has focused on compact star modeling, exploring stellar structure, mass-radius relations, and stability bounds in $f(\Q)$ gravity \cite{nashed2024structuremaximummassstability,sharma2024physicalpropertiesmaximumcompactness,dearaujo2024compactstarsfq,Maurya_2024,Maurya_2022,Lin_2021,Das_2024zsg,Alwan_2024,maurya2023effectgravitationaldecouplingconstraining,Pradhan_2024,10.1093mnrasstaf1999}. Also most of these studies adopt either the coincident gauge or a spherical GR-like connection, so that potential beyond-GR signatures associated with the affine structure are largely fixed rather than followed dynamically. For beyond stellar configurations, similar simplifying strategies are often employed in gravastar studies \cite{Sharif_2025,ibrar2025analyzinggravastarstructurefinchskea,Mohanty_2024,Mohanty_2024_2,Javed_2024,Pradhan_2023,Pradhan_2023_2,bhattacharjee2025} and more general static spherically symmetric solutions \cite{awais2025anisotropiccompactstarvaidyatikekar,Maurya2023,Ditta2023,Maurya_2022vsn,Calz__2023,Bhar_2024,Sokoliuk_2022,PhysRevD.105.024060,Bhar_2023,Iqbal:2025iln,paul2025studyphysicalpropertiescharacteristics,das2025studypulsarexo1745248,sharif2025chargedanisotropicpulsarsax,sharif2025impactpulsarsaxj174892021}. Complementary to these approaches, other works do place particular emphasis on the connection as a dynamical object, with some addressing this issue in cosmological settings \cite{DAmbrosio2021,Narawade_2025,abebe2025noncoincidencefq,de2025exactcosmo,ayuso2025insightsfqcosmologyrelevance} and others developing more general frameworks \cite{Bahamonde_2022} or studying compact objects directly \cite{dimakis2025relativisticstarsfqgravity,dimakis2024staticsphericallysymmetricsolutions,Bahamonde_2022_2}. Taken together, this expanding body of work signals a healthy and growing theoretical interest in $f(\Q)$ gravity, but the diversity in how the connection is treated can make it challenging to assess the theory's full implications in a unified way. In this context, a careful and systematic analysis —such as the one we present here for NS—, in which we aim to complement and extend existing studies by exploring NS configurations in which the connection is consistently treated as an active, dynamical ingredient, becomes not only timely but necessary.

In this work, we lay the theoretical and numerical groundwork necessary to explore beyond-GR NS solutions in $f(\Q)$ gravity. Our primary aim is to clarify under what conditions non-trivial solutions can exist and to identify the assumptions that must be avoided to preserve the theory’s additional degrees of freedom. Because these solutions typically lie beyond analytic reach, we analyze the regularity of the system to formulate a numerically solvable BVP required for consistent integration. This includes determining appropriate boundary conditions, the form of the metric and connection near the center, and the expected asymptotic behavior. This process clarifies how the connection’s degrees of freedom naturally constrain the admissible form of viable beyond-GR solutions and help us identify the numerical obstacles that must be overcome. We then examine the specific numerical challenges that arise when solving the equations under these more general conditions and outline strategies for addressing them. Our focus here is thus methodological. We aim to establish a robust framework for future studies seeking to go beyond GR in a controlled and reliable way, while leaving the numerical construction of complete stellar models for future work.

This paper is organized as follows. In Sec. \ref{sec:GeneralIntro}, we briefly review the geometric structure of STEGR and its generalization to the framework in $f(\Q)$ gravity, introducing the essential elements needed to define our conventions and set up the field equations. Sec. \ref{sec:SphSols} concisely establishes the general structure for static and stationary spherically symmetric configurations, including the most general flat, torsionless connection compatible with these symmetries —previously derived for BHs in \cite{DAmbrosio2022}— and featuring one potentially dynamical, non-trivial component. We present the corresponding equations of motion adapted in the presence of matter and in Sec. \ref{sec:SphSols_PerfectFluid} and Sec. \ref{sec:ChooseEoS}, we introduce a convenient metric reparameterization and outline the approximations used in the fluid description adopted for NS. Sec. \ref{sec:STEGRlim} 
focuses on identifying the assumptions that reduce the theory to GR, either through specific limits of the function $f(\Q)$ or by prematurely fixing the connection. This step is important for the numerical analysis that follows, as it provides a controlled GR limit from which to depart in the search for beyond-GR solutions. We show that even mild conditions on the connection components can suppress the theory’s additional degrees of freedom and thus prevent genuinely new behavior from emerging. Sec. \ref{sec:beyondGR} explores beyond-GR NS solutions for two representative $f(\Q)$ models: $f(\Q) = \Q + \alpha \Q^2$ and $f(\Q) = \Q^\beta$. We begin in Sec. \ref{sec:beyondGR_constraints} by analyzing the regularity and boundary behavior of the field equations at the center of the NS and at large distances, establishing consistent boundary conditions required for numerical integration. Through a perturbative analysis, we show that standard expansion techniques yield GR-like solutions. Specifically, solutions admitting a series expansion near $r=0$, or in powers of $1/r$ at far infinity, seem to converge to GR behavior. This constrains the asymptotic structure that non-trivial solutions may exhibit, guiding the numerical strategy developed in Sec. \ref{sec:beyondGR_numerics}. There, we identify and address the key challenges in solving the full system under general conditions. We do not present numerical stellar models, but provide instead the boundary data, consistency checks, and continuation road map that enable such computations. Finally, Sec. \ref{sec:conclusions} summarizes our findings and highlights potential directions for future research.

\vspace{-1.4em}
\section{A description of gravity based on non-metricity}
\label{sec:GeneralIntro}

\subsection{Symmetric Teleparallelism}
\label{sec:GeneralIntro_STEGR}

Consider a metric-affine geometry, described by the triplet $(\mathcal{M}, g_{\mu\nu}, \Gamma^{\alpha}_{\mu\nu})$, where $\mathcal{M}$ is a smooth four-dimensional manifold, $g_{\mu\nu}$ are the components of a metric tensor with signature $(-,+,+,+)$, and $\Gamma^{\alpha}_{\mu\nu}$ represents the affine connection. The latter defines how vectors are parallel transported along curves and establishes the form of the covariant derivative:
\begin{equation}
	\nabla_{\mu}u^{\rho}=\partial_{\mu}u^{\rho}+\Chr{\rho}{\mu}{\nu}u^{\nu}.
	\label{eq:CovDev}
\end{equation}
The choice of connection enables the construction of tensors that quantify the geometric behavior of the manifold, namely the Riemann curvature tensor $R\idx{^\rho_{\alpha\mu\nu}}$, the torsion tensor $T\idxudd{\alpha}{\mu}{\nu}$, and the non-metricity tensor $Q\idx{_{\alpha\mu\nu}}$, which together offer three complementary approaches to gravity \cite{Heisenberg_2019, GeomTrinity}.

The first two are defined as:
\begin{align}
	R&\idx{^\rho_{\alpha\mu\nu}}\coloneqq\DChr{\rho}{\nu}{\alpha}{\mu}-\DChr{\rho}{\mu}{\alpha}{\nu}+\Chr{\rho}{\mu}{\beta}\Chr{\beta}{\nu}{\alpha}-\Chr{\rho}{\nu}{\beta}\Chr{\beta}{\mu}{\alpha},\\
    T&\idxudd{\alpha}{\mu}{\nu}\coloneqq\Chr{\alpha}{\mu}{\nu}-\Chr{\alpha}{\nu}{\mu};
\end{align}
and are set to zero for a description of gravity based entirely on non-metricity. The third and only non-trivial tensor characterizes the change in length of a vector as it is parallel transported and is defined as:
\begin{equation}
Q\idx{_{\alpha\mu\nu}}\coloneqq\nabla_{\alpha}g_{\mu\nu}=\partial_{\alpha}g_{\mu\nu}-2\Gamma^{\lambda}_{\alpha(\mu}g_{\nu)\lambda},
\label{eq:NonMetricityTensor}
\end{equation}
being thus symmetric in its last two indices. Because of this property, there are two independent traces that can be constructed, namely $Q_{\alpha}\coloneqq Q\idx{_{\alpha}^{\lambda}_{\lambda}}$ and $\tilde{Q}_{\alpha}\coloneqq Q\idx{_{\lambda\alpha}^{\lambda}}$, which in turn give rise to five independent contractions at the quadratic order in non-metricity. The most general combination of these five contractions results in the definition of the non-metricity scalar $\Q$ \cite{Heisenberg_2019, GeomTrinity, Jim_nez_2018, Jim_nez_2018_2}:
\begin{align}
		\Q &\coloneqq c_1 Q_{\alpha\beta\gamma}Q^{\alpha\beta\gamma}
		+c_2Q_{\alpha\beta\gamma}Q^{\beta\alpha\gamma}+c_3Q_{\alpha}Q^{\alpha}\nonumber\\ 
		&+c_4 \tilde{Q}_{\alpha}\tilde{Q}^{\alpha}+c_5Q_{\alpha}\tilde{Q}^{\alpha}.
	\label{eq:NonMetricityScalar}
\end{align}
For later convenience, we introduce the \textit{non-metricity conjugate} $P\idxudd{\alpha}{\mu}{\nu}$ and the symmetric tensor $q_{\mu\nu}$. The former is defined as
\begin{align}
		P\idxudd{\alpha}{\mu}{\nu} &\coloneqq\,\frac{1}{2}\frac{\partial\Q}{\partial Q\idx{_{\alpha}^{\mu\nu}}}=c_1\, Q\idx{^{\alpha}_{\mu\nu}}+c_2\,Q\idx{_{(\mu}^{\alpha}_{\nu)}}+c_3\,g_{\mu\nu}Q^{\alpha}\nonumber\\
		&+c_4\,\delta^{\alpha}\, _{(\mu}\tilde{Q}_{\nu)}+\frac{1}{2}c_5\,\left(g_{\mu\nu}\tilde{Q}^{\alpha}+\delta^{\alpha}\, _{(\mu}Q_{\nu)}\right),
	\label{eq:NonMetricityConjugate}
\end{align}
it is manifestly symmetric in its lower two indices and it fulfills the property
\begin{equation}
		\Q=P_{\alpha\mu\nu}Q^{\alpha\mu\nu}.
\end{equation}
$q_{\mu\nu}$, on the other hand, is defined as:
\begin{align}
		q_{\mu\nu}&\coloneqq\,\frac{\partial\Q}{\partial g^{\mu\nu}}=c_1\,(Q\idx{_{\mu}^{\alpha\beta}}Q_{\nu\alpha\beta}-2Q\idx{_{\alpha\mu}^{\beta}}Q\idxudd{\alpha}{\nu}{\beta})\nonumber\\
		&-c_2\,Q\idx{_{\alpha\mu}^{\beta}}Q\idx{_{\beta\nu}^{\alpha}}+c_3\, (Q_{\mu}Q_{\nu}-2Q_{\alpha\mu\nu}Q^{\alpha})\nonumber\\
		&-c_4\,\tilde{Q}_{\mu}\tilde{Q}_{\nu}-c_5\,Q_{\alpha\mu\nu}\tilde{Q}^{\alpha},
\end{align}
and we can rewrite it in a more compact and manifestly symmetric form making use of the non-metricity conjugate as
\begin{equation}
		q_{\mu\nu}=P_{(\mu|\alpha\beta}Q_{\nu)}\, ^{\alpha\beta}-2P^{\alpha\beta}\,_{(\nu}Q_{\alpha\beta|\mu)}.
	\label{eq:SymmetricTensor_q}
\end{equation}
The most general action for a flat, torsion-free, non-metric theory with quadratic EoMs reads then \cite{Heisenberg_2019, GeomTrinity, Jim_nez_2018}:
\begin{align}
	\mathcal{S}_{\Q}[g,\Gamma;\lambda,\rho]\coloneqq\int_{\mathcal{M}}d^4x\bigg(&\frac{1}{2\kappa}\sqrt{-g}\,\Q+\lambda\idx{_{\alpha}^{\beta\mu\nu}}R\idx{^{\alpha}_{\beta\mu\nu}}\nonumber\\
	&+\rho\idxudd{\alpha}{\mu}{\nu}T\idxudd{\alpha}{\mu}{\nu}\bigg)+\mathcal{S}_{\rm matter},
	\label{eq:ActionQGeneric}
\end{align}
where $\kappa=8\pi$ in geometrized units and the factor $\sqrt{-g}$ ensures that our volume element $\sqrt{-g}\,d^4x$ is invariant under coordinate transformations as in the Einstein-Hilbert action in GR. The conditions of flatness and no torsion are imposed at the level of the action by performing variations with respect to the two Lagrange multiplier fields $\lambda\idx{_{\alpha}^{\beta\mu\nu}}=\lambda\idx{_{\alpha}^{\beta[\mu\nu]}}$ and $\rho\idx{_{\alpha}^{\mu\nu}}=\rho\idx{_{\alpha}^{[\mu\nu]}}$, defined as antisymmetric tensor densities in their last two indices \cite{Jim_nez_2018, Jim_nez_2018_2}. By solving the constraints, it can be shown that the flatness and vanishing torsion conditions restrict the connection to be purely inertial so that, for arbitrary $\xi^{\alpha}$, it becomes \cite{Heisenberg_2019, GeomTrinity, Jim_nez_2018, Jim_nez_2018_2, DAmbrosio2022}:
\begin{equation}
	\Chr{\alpha}{\mu}{\nu}=\frac{\partial x^{\alpha}}{\partial \xi^{\lambda}}\partial_{\mu}\partial_{\nu}\xi^{\lambda}.
	\label{eq:ConnectionGaugeFreedom}
\end{equation}
The affine structure of the theory reveals thus one of the characteristic properties of symmetric teleparallelism: the connection can be arbitrarily chosen and there exists a special gauge choice, the so-called \textit{coincident gauge}, in which $\xi^{\alpha}=x^{\alpha}$ and the connection becomes trivial, i.e., $\Chr{\alpha}{\mu}{\nu}=0$.  Mathematically speaking, this means that all covariant derivatives are replaced by ordinary derivatives. Physically speaking, the connection becomes inertial and the gauge itself can be interpreted as the one where the origin of the tangent space, parametrized by $\xi^{\alpha}$, coincides with the spacetime origin, hence the name \cite{Jim_nez_2018, Heisenberg_2019}.

To recover the STEGR theory, one starts by re-writing a generic connection as:
\begin{equation}
	\Chr{\alpha}{\mu}{\nu}=\left\{\begin{smallmatrix}\alpha \\ \mu\nu\end{smallmatrix}\right\}+K\idxudd{\alpha}{\mu}{\nu}(T)+L\idxudd{\alpha}{\mu}{\nu}(Q),
	\label{eq:GeneralConnection3terms}
\end{equation}
where $\left\{\begin{smallmatrix}\alpha \\ \mu\nu\end{smallmatrix}\right\}$ stands for its Levi-Civita part, $K\idxudd{\alpha}{\mu}{\nu}\coloneqq \frac{1}{2}T\idxudd{\alpha}{\mu}{\nu}+T\idx{_{(\mu}^{\alpha}_{\nu)}}$ is the so-called \textit{contorsion tensor} and carries the contributions coming from the torsion, and the \textit{disformation tensor} $L\idxudd{\alpha}{\mu}{\nu}\coloneqq \frac{1}{2}Q\idxudd{\alpha}{\mu}{\nu}-Q\idx{_{(\mu}^{\alpha}_{\nu)}}$ gathers the contributions of the non-metricity \cite{Heisenberg_2019, GeomTrinity, Jim_nez_2018, Jim_nez_2018_2, R_nkla_2018}. This last two pieces are sometimes referred to as \textit{distortion}.

Under such a shift of the connection, two contractions of the corresponding expansion of the Riemann tensor, together with the two geometrical postulates of symmetric teleparallelism, $T\idxudd{\alpha}{\mu}{\nu}\stackrel{!}{=}0$, $R\idx{^\rho_{\alpha\mu\nu}}\stackrel{!}{=}0$, lead to the following relation between the Ricci scalar for the Levi-Civita connection, $\mathcal{R}$, and the non-metricity scalar \cite{Heisenberg_2019,GeomTrinity, Jim_nez_2018_2, DAmbrosio2022}:
\begin{equation}
	\mathcal{R}=-\mathring{\Q}-\mathcal{D}_{\alpha}(Q^{\alpha}-\tilde{Q}^{\alpha}).
	\label{eq:EqbetweenRicciAndQ2}
\end{equation}
Here, $\mathring{\Q}$ stands for the non-metricity scalar $\Q$ in (\ref{eq:NonMetricityScalar}) for the specific choice of coefficients $\{c_1,c_2,c_3,c_4,c_5\}=\left\{-1/4,1/2,1/4,0,-1/2\right\}$. From this expression it becomes evident that the action
\begin{equation}
	\mathcal{S}_{\rm STEGR}[g,\Gamma]=-\frac{1}{2\kappa}\int_{\mathcal{M}}d^4x\sqrt{-g}\, \mathring{\Q}
\end{equation}
differs from the Einstein-Hilbert action only by a total derivative, reproducing thus the dynamics and phenomenology of GR. In addition to having a description in which the curvature tensor with respect to the affine connection vanishes, (\ref{eq:EqbetweenRicciAndQ2}) shows that we are also subtracting the derivatives in the boundary term from the usual action in GR. These represent inertial effects that typically diverge away from the source and consequently, in a Teleparallel theory of gravity the covariance of the theory is not compromised \cite{Heisenberg_2019, GeomTrinity, Jim_nez_2018_2}.

\subsection{General framework in $f(\Q)$ gravity}
\label{sec:GeneralIntro_fQ}

The leap from (\ref{eq:ActionQGeneric}) to $f(\Q)$ gravity \cite{heisenberg2023reviewfqgravity} is trivial by promoting the non-metricity scalar to an arbitrary function $f(\mathring{\Q})$, since only our choice of $\{c_i\}_{i=1}^5$ contains the STEGR.  In order to alleviate the notation, however, the symbol $\Q$ is used henceforth instead. In this sense, the most general action in $f(\Q)$ gravity, in some convenient system of units reads:
\begin{align}
	\mathcal{S}_{f(\Q)}[g,\Gamma;\lambda,\rho]\coloneqq\int_{\mathcal{M}}&d^4x\bigg(\frac{1}{2}\sqrt{-g}\,f(\Q)+\lambda\idx{_{\alpha}^{\beta\mu\nu}}R\idx{^{\alpha}_{\beta\mu\nu}}\nonumber\\
	&+\rho\idxudd{\alpha}{\mu}{\nu}T\idxudd{\alpha}{\mu}{\nu}\bigg)+\mathcal{S}_{\rm matter}.
	\label{eq:ActionfQGeneric}
\end{align}
The EoMs of the theory are computed in their most generic form and in a manifestly covariant formulation following an approach à la Palatini and performing a variation of the action functional with respect to the metric and with respect to the connection. Note that contrary to what happens in STEGR, the latter is not entirely determined by the metric and can contribute additional degrees of freedom in this case, as detailed in \ref{sec:SphSols_SymReduction}. The matter distribution under consideration is encoded in $\mathcal{S}_{\rm matter}$, whose variation with respect to the metric and connection results in the usual definition for the energy-momentum tensor and hypermomentum, respectively \cite{Jim_nez_2018,Jrv2018,Jim_nez_2018_2,R_nkla_2018}:
\begin{equation}
	\mathcal{T}_{\mu\nu}\coloneqq -\frac{2}{\sqrt{-g}}\frac{\delta \mathcal{S}_{\rm matter}}{\delta g^{\mu\nu}}, \quad \varvarfrakh\idx{_{\alpha}^{\mu\nu}}\coloneqq -\frac{1}{2}\frac{\delta \mathcal{S}_{\rm matter}}{\delta\Chr{\alpha}{\mu}{\nu}}.
\end{equation}
$\varvarfrakh\idx{_{\alpha}^{\mu\nu}}$ is generated by coupling fermions to gravitation and its presence or absence is therefore determined by the way we choose to couple the matter fields in our framework \cite{Jim_nez_2018_2, BeltranJimenez:2020sih}. Despite the ambiguity in such choice for generalized geometries, from the evidence we have from GR, we are interested in considering situations in which there is absence of any coupling to the connection. Thus, in the following we always consider $\varvarfrakh\idx{_{\alpha}^{\mu\nu}}=0$.

The stationary-action principle on the variation of (\ref{eq:ActionfQGeneric}) with respect to the inverse of the metric tensor $g^{\mu\nu}$, $\delta_{g}\mathcal{S}_{f(\Q)}\stackrel{!}{=} 0$, yields the EoMs for the metric \cite{Jim_nez_2018, Jim_nez_2018_2,R_nkla_2018, Jim_nez_2020, Zhao_2022,Lin_2021,DAmbrosio2022, DAmbrosio2021}:
\begin{align}
	\mathcal{M}_{\mu\nu}&\coloneqq\frac{2}{\sqrt{-g}}\nabla_{\alpha}\left[\sqrt{-g}\,P\idxudd{\alpha}{\mu}{\nu}\fQ\right]+\fQ\, q_{\mu\nu}\nonumber\\
	&-\frac{1}{2}f{(\Q)}g_{\mu\nu}=\kappa\mathcal{T}_{\mu\nu},
	\label{eq:EoMMetricCovariant}
\end{align}
where we are using the prime notation to denote the derivatives of $f$ with respect to $\Q$, namely $\fQ\coloneqq df(\Q)/d\Q$.
These equations can be brought into the more practical and suggestive form \cite{DAmbrosio2022, Zhao_2022}
\begin{align}
	\mathcal{M}_{\mu\nu}&=\fQ \mathcal{G}_{\mu\nu}-\frac{1}{2}g_{\mu\nu}\left[f(\Q)-\fQ\Q\right]\nonumber\\
	&+2\fQQ P\idxudd{\alpha}{\mu}{\nu}(\partial_{\alpha}\Q)=\kappa\mathcal{T}_{\mu\nu},
	\label{eq:EoMMetricCovariant2}
\end{align}
where $\mathcal{G}_{\mu\nu}\coloneqq\mathcal{R}_{\mu\nu}-(1/2)\mathcal{R}g_{\mu\nu}$ is the Einstein tensor (computed with respect to the Levi-Civita connection). This way of writing the EoMs is more transparent and practical in the sense that it prevents one from calculating a covariant derivative and evidences the fact that the choice $f(\Q)=\Q$ recovers the STEGR limit of the theory. For this case, it follows that $\mathcal{G}_{\mu\nu}=\mathcal{T}_{\mu\nu}$ and we thus recover the same dynamics as in standard GR.

The connection EoMs are derived by performing a variation with respect to the connection, $\delta_{\Gamma}\mathcal{S}_{f(\Q)}\stackrel{!}{=} 0$, obtaining \cite{Jim_nez_2018, Jim_nez_2018_2}:
\begin{equation}
	\nabla_{\beta}\lambda\idx{_{\alpha}^{\nu\mu\beta}}+\rho\idx{_{\alpha}^{\mu\nu}}=\sqrt{-g}\fQ P\idx{^{\mu\nu}_{\alpha}},
	\label{eq:ConnectionEoMPreliminar}
\end{equation}
where one assumes, also in the derivation of (\ref{eq:EoMMetricCovariant}), that the variations of the metric tensor and connection vanish on the boundary, i.e., $\delta g^{\mu\nu}|_{\partial \mathcal{M}}=\delta\Chr{\alpha}{\mu}{\nu}|_{\partial\mathcal{M}}=0$. Two further covariant derivatives acting on (\ref{eq:ConnectionEoMPreliminar}) remove the Lagrange multipliers due to their symmetry properties and the commutation of covariant derivatives in STG ---since the commutator of a general tensor field is proportional to curvature and torsion terms---, leading to the more compact form \cite{Jim_nez_2018_2,R_nkla_2018,Jim_nez_2020,DAmbrosio2022,Zhao_2022}:
\begin{equation}
\mathcal{C}_{\alpha}\coloneqq\nabla_{\mu}\nabla_{\nu}\left(\sqrt{-g}\,\fQ P\idx{^{\mu\nu}_{\alpha}}\right)=0.
\label{eq:EoMConnectionCovariant}
\end{equation}
The expansion of (\ref{eq:EoMConnectionCovariant}) to find their explicit form in terms of the unknown components of a certain metric and connection requires the computation of two covariant derivatives, also with respect to such arbitrary connection that can potentially become dynamical. Bianchi identities turn out to be a convenient tool that offer a much more efficient alternative route to circumvent the cumbersome calculation. These identities take the form, for $f(\Q)$ gravity:
\begin{equation}
\mathring{\nabla}_{\mu}\mathcal{M}\idx{^{\mu}_{\nu}}+\mathcal{C}_{\nu}=0;
\label{eq:BianchiIdentities}
\end{equation}
where the operator $\mathring{\nabla}$ denotes the covariant derivative with respect to the Levi-Civita connection. A formal definition can be found in \cite{heisenberg2023reviewfqgravity}. Intuitively, they can be understood as a consequence of imposing the conservation of the energy-momentum tensor, $\mathring{\nabla}^{\mu}\mathcal{T}_{\mu\nu}=0$, on (\ref{eq:EoMMetricCovariant}) or (\ref{eq:EoMMetricCovariant2}), so that $\mathring{\nabla}^{\mu}\mathcal{M}_{\mu\nu}=0$. Since from (\ref{eq:EoMConnectionCovariant}), $\mathcal{C}_{\nu}=0$, then these two results imply (\ref{eq:BianchiIdentities}). Therefore, a faster way to obtain the connection EoMs, is to take the covariant derivative of the metric ones 
 with respect to the Levi-Civita connection and set it to be equal to zero.

\section{Stationary and spherically symmetric neutron star solutions}
\label{sec:SphSols}
Let us recall that the focus of this work is to investigate NS solutions within this formalism, and in particular to do so without imposing too restrictive assumptions that kill some of the degrees of freedom of the theory. To this end, one needs to enforce on both the EoMs and the connection the appropriate symmetries for our scenario: stationary spacetimes with spherical symmetry. Since unlike in GR, the latter implies considering what connection is compatible not only with the symmetries of our spacetime but also with the postulates of symmetric teleparallelism.  Once these symmetry constraints are implemented, we can proceed to specify our distribution of energy and momentum (i.e., our $\mathcal{T}_{\mu\nu}$), and if necessary, to close the system of equations to be solved. In the case of NS solutions, this also involves defining an EoS.

\subsection{Metric, connection and EoMs}
\label{sec:SphSols_SymReduction}

We work in spherical coordinates, i.e., we assume that both the components of the metric and the affine connection can be expressed in the chart $(t,r,\theta,\phi)\in\mathbb{R}\times\mathbb{R}_{>0}\times[0,\pi]\times[0,2\pi)$. For the former, we choose to employ the simplest form that the metric tensor of a stationary and spherically symmetric geometry can take. This is, without loss of generality, given by
%
%
\begin{equation}
g_{\mu\nu} = \begin{pmatrix}
g_{tt} & 0 & 0 & 0 \\
0 & g_{rr} & 0 & 0 \\
0 & 0 & r^2 & 0 \\
0 & 0 & 0 & r^2\sin^2{\theta} \\
\end{pmatrix},
\label{eq:ReducedMetric}
\end{equation}
where $g_{tt}=g_{tt}(r)$ and $g_{rr}=g_{rr}(r)$ only depend on $r$ \cite{DAmbrosio2022,misner2017gravitation,schutz2009first}, having thus reduced the number of independent components to be determined from 10 to 2.
The symmetry reduction for the connection, however, is significantly more complicated and has been done in full detail in \cite{DAmbrosio2022}. In this regard, we refer to this study for more details and present in TABLE \ref{Tab:ConnectionComponents} the components of the flat, torsionless, stationary and spherically symmetric connection that we consider for our calculations.

\begin{table*}[hbtp!]
  \centering 
  \caption{A concise summary of all the components and properties of the flat, torsionless, stationary and spherically symmetric connection we use in this work. Table and caption adapted from Table 2 and Table 3 in \cite{DAmbrosio2022}.}
    \label{Tab:ConnectionComponents}
    \vspace{5pt}
   \renewcommand{\arraystretch}{1.5}
    \begin{tabular}{|p{3cm}||p{12.9cm}|}
	\hline
	Independent\newline components & All connection components can be expressed in terms of an arbitrary constant $\Chr{t}{\theta}{\theta}$, the two functions $\Chr{r}{r}{r}(r),\Chr{r}{\theta}{\theta}(r)$, with $\Chr{r}{\theta}{\theta}(r)\neq 0$; and trigonometric functions. \\
    \hline
      Non-zero\newline components & {There are 10 non-zero components: the three independent components \newline 
      $\Chr{t}{\theta}{\theta}=\mathrm{const.}$, $\Chr{r}{r}{r}(r)$, $\Chr{r}{\theta}{\theta}(r)$ and \vspace{0.1cm} \newline 
      $\begin{aligned} 
     &\Chr{t}{r}{r}=-\tfrac{\Chr{t}{\theta}{\theta}}{(\Chr{r}{\theta}{\theta})^2}, \qquad\qquad\:\:\:\Chr{\theta}{r}{\theta}=-\tfrac{1}{\Chr{r}{\theta}{\theta}}, \qquad\qquad\quad\Chr{\phi}{r}{\phi}=-\tfrac{1}{\Chr{r}{\theta}{\theta}}, \\
     &\Chr{t}{\phi}{\phi}=\sin^2{\theta}\,\Chr{t}{\theta}{\theta},\qquad\quad\;\!\:\:\!\:\:\!\Chr{\theta}{\phi}{\phi}=-\cos{\theta}\sin{\theta}, \quad\quad\:\!\!\Chr{\phi}{\theta}{\phi}=\cot{\theta},\\
     &\Chr{r}{\phi}{\phi}=\sin^2{\theta}\,\Chr{r}{\theta}{\theta}.
        \end{aligned}$\vspace{0.2cm}}\\
   \hline
   Derivatives of\newline independent\newline components & {Of the two independent functions, the $r$-derivative of $\Chr{r}{\theta}{\theta}$ can be expressed as \vspace{0.1cm} \newline $\partial_r\Chr{r}{\theta}{\theta}=-1-\Chr{r}{r}{r}\Chr{r}{\theta}{\theta}$, \vspace{0.1cm} \newline while $\partial_r\Chr{r}{r}{r}$ cannot be expressed in terms of other components.} \\
   \hline
  \end{tabular}
  \vspace{-7pt}
\end{table*}

In particular, note that all components can be expressed in terms of arbitrary constants, trigonometric functions or other components, except for $\Chr{r}{r}{r}(r)$ and $\Chr{r}{\theta}{\theta}(r)$. Since there exists only one differential equation for $\partial_r\Chr{r}{\theta}{\theta}$, the background connection has thus one non-trivial component that can potentially become dynamical. Also, the coincident gauge (see equation (\ref{eq:ConnectionGaugeFreedom})) fails to be spherically symmetric and therefore any attempt to find such solutions to the field equations of $f(\Q)$ gravity, while keeping the simple form of the metric tensor depicted in (\ref{eq:ReducedMetric}), is doomed to failure~\cite{DAmbrosio2022}. For later convenience, we also stress that, given that the derivative equation only involves $\Gamma^r_{\theta\theta}(r)$ and $\Gamma^r_{rr}(r)$, we can easily trade one for the other and decide which one to treat as our independent connection variable. Here and in what follows, we use the compact notation $\partial_r X$ for radial derivatives; since all quantities depend only on $r$, this should be understood as a total derivative $dX/dr$.

Likewise, we can solve the explicit expression for the non-metricity scalar for $\Chr{r}{r}{r}$.

\begin{align}
\Q&=\frac{1}{r^2 g_{rr}^2 g_{tt} (\Chr{r}{\theta}{\theta})^2} \big\{g_{tt} \big[\Chr{r}{\theta}{\theta} \big((\Chr{r}{\theta}{\theta})^2g_{rr}-r^2\big) (\partial_r g_{rr})\nonumber\\
&+2 g_{rr} \big((\Chr{r}{\theta}{\theta})^2-g_{rr}\Chr{r}{r}{r}(\Chr{r}{\theta}{\theta})^3+r^2+r \Chr{r}{\theta}{\theta} (r \Chr{r}{r}{r}+2)\big)\big]\nonumber\\
&+\Chr{r}{\theta}{\theta}g_{rr} \big[(\Chr{r}{\theta}{\theta})^2g_{rr}+2 r \Chr{r}{\theta}{\theta}+r^2\big] (\partial_r g_{tt})\big\},
\label{eq:NonMetricityScalarConnection}
\end{align}
and use $\Q$ instead. This should be seen as a measure taken for simplicity and practical reasons.

Concerning the field equations (\ref{eq:EoMMetricCovariant}) and (\ref{eq:EoMConnectionCovariant}), the enforcing of the symmetries reduces their structure to only three non-trivial components: $\mathcal{M}_{tt}$, $\mathcal{M}_{rr}$ and $\mathcal{C}_r$. We refer again to \cite{DAmbrosio2022} for the details. In this sense, our theory needs to be solved for either $\{g_{tt}, g_{rr}, \mathbb{Q}\}$, $\{g_{tt}, g_{rr}, \Gamma^r_{rr}\}$ or $\{g_{tt}, g_{rr}, \Gamma^r_{\theta\theta}\}$ ---we stick to the first scenario for simplicity--- for such three EoMs. To obtain their explicit form, one solves (\ref{eq:NonMetricityScalarConnection}) for $\Chr{r}{r}{r}$ to trade these two variables and use $\Q$ in the rest of the necessary equations instead. Next one calculates the EoMs $\mathcal{M}_{tt}$ and $\mathcal{M}_{rr}$ using (\ref{eq:EoMMetricCovariant2}) and solve $\mathcal{M}_{rr}$ for $\partial_r g_{tt}$ and $\mathcal{M}_{tt}$ for $\partial_r g_{rr}$, obtaining:
\begin{align}
\partial_r g_{tt} = &+\frac{rg_{tt}\kappa}{f'(\mathbb{Q})}\mathcal{T}_{rr}-\frac{g_{tt}\big[2+g_{rr}\big(\mathbb{Q}r^2-2-\frac{f(\mathbb{Q})r^2}{f'(\mathbb{Q})}\big)\big]}{2r}\nonumber\\
&+\frac{g_{tt}\big[r^2-g_{rr}(\Gamma^r_{\theta\theta})^2\big]}{\Gamma^r_{\theta\theta} r f'(\mathbb{Q})}(\partial_r\mathbb{Q})f''(\mathbb{Q}),\label{eq:EoMMetricgttReduced}
\end{align}
\begin{align}
\partial_r g_{rr} = &-\frac{rg_{rr}^2\kappa}{f'(\mathbb{Q})g_{tt}}\mathcal{T}_{tt}+\frac{g_{rr}\big[2+g_{rr}\big(\mathbb{Q}r^2-2-\frac{f(\mathbb{Q})r^2}{f'(\mathbb{Q})}\big)\big]}{2r}\nonumber\\
&+\frac{g_{rr}\big[(\Gamma^r_{\theta\theta})^2g_{rr}+2r\Gamma^r_{\theta\theta}+r^2\big]}{\Gamma^r_{\theta\theta} r f'(\mathbb{Q})}(\partial_r\mathbb{Q})f''(\mathbb{Q}).
\label{eq:EoMMetricgrrReduced}
\end{align}
These equations identically correspond to those obtained in \cite{DAmbrosio2022} for BH solutions, now including an arbitrary (stationary and spherically symmetric) energy-momentum distribution encoded in $\mathcal{T}_{\mu\nu}$.

%
%
%
%

To ensure that the only derivatives of the metric components appear in (\ref{eq:EoMMetricgttReduced}) and (\ref{eq:EoMMetricgrrReduced}), we can further use these two equations to substitute those appearing in the result of solving $\Q$ in (\ref{eq:NonMetricityScalarConnection}) for $\Chr{r}{r}{r}$. This also allows us to write down the equation for the derivative $\partial_r\Chr{r}{\theta}{\theta}$ appearing in the last row of TABLE \ref{Tab:ConnectionComponents}. These two utility equations are:
\begin{widetext}
\begin{align}
\Gamma^r_{rr}&=\frac{\Gamma^r_{\theta\theta}f(\mathbb{Q})g_{rr}r^2(\Gamma^r_{\theta\theta}+r)+\big\{2r^2+\Gamma^r_{\theta\theta}\big[2r+g_{rr}\big(2(r+\Gamma^r_{\theta\theta})-r^2\mathbb{Q}(r+2\Gamma^r_{\theta\theta})\big)\big]\big\}f'(\mathbb{Q})}{2f'(\mathbb{Q})\Gamma^r_{\theta\theta}\big[(\Gamma^r_{\theta\theta})^2g_{rr}-r^2\big]}\nonumber\\
&-\frac{r\kappa}{2f'(\mathbb{Q})}\Bigg\{\frac{g_{rr}}{g_{tt}}\mathcal{T}_{tt}-\frac{(\Gamma^r_{\theta\theta})^2g_{rr}+2r\Gamma^r_{\theta\theta}+r^2}{(\Gamma^r_{\theta\theta})^2g_{rr}-r^2}\mathcal{T}_{rr}\Bigg\},\label{eq:QSolvedForGammarrr}
\end{align}
\begin{align}
\partial_r\Gamma^r_{\theta\theta}&=-\frac{\Gamma^r_{\theta\theta}\big\{f(\mathbb{Q})g_{rr}r^2(\Gamma^r_{\theta\theta}+r)+\big[2r-g_{rr}(2\Gamma^r_{\theta\theta}+r)(\mathbb{Q}r^2-2)\big]f'(\mathbb{Q})\big\}}{2f'(\mathbb{Q})\big[(\Gamma^r_{\theta\theta})^2g_{rr}-r^2\big]}\nonumber\\
&+\frac{r\kappa\Gamma^r_{\theta\theta}}{2f'(\mathbb{Q})}\Bigg\{\frac{g_{rr}}{g_{tt}}\mathcal{T}_{tt}-\frac{(\Gamma^r_{\theta\theta})^2g_{rr}+2r\Gamma^r_{\theta\theta}+r^2}{(\Gamma^r_{\theta\theta})^2g_{rr}-r^2}\mathcal{T}_{rr}\Bigg\}.\label{eq:ConstraintSolvedrthetatheta}
\end{align}
\end{widetext}
Note that (\ref{eq:QSolvedForGammarrr}) does not provide any relevant dynamical information nor shall it be used to solve our system, but it is the one that allows us to trade $\Chr{r}{r}{r}\leftrightarrow\Q$ and written in this way, it already depends on the variables we want to use; i.e., it does not introduce any additional derivative of the metric when making such replacement. (\ref{eq:ConstraintSolvedrthetatheta}) on its side constitutes a constraint equation for this non-dynamical component.

Finally, by acting with the covariant derivative $\mathring{\nabla}_{\mu}$ on (\ref{eq:EoMMetricCovariant2}) (with the corresponding upper index), according to the Bianchi identities (\ref{eq:BianchiIdentities}), and solving for $\partial^2_r\Q$; we can easily calculate the EoM of the connection $\mathcal{C}_r$, obtaining:
\begin{widetext}
\begin{align}
\partial^2_r\Q =&-\frac{r\kappa(\partial_r\mathbb{Q})}{2f'(\mathbb{Q})}\Bigg\{\mathcal{T}_{rr}-\frac{g_{rr}\big[(\Gamma^r_{\theta\theta})^2g_{rr}+r^2\big]}{g_{tt}\big[(\Gamma^r_{\theta\theta})^2g_{rr}-r^2\big]}\mathcal{T}_{tt}\Bigg\}+\frac{(\partial_r\mathbb{Q})}{2\big[(\Gamma^r_{\theta\theta})^2g_{rr}-r^2\big]}\Bigg\{2r+g_{rr}\big[4\Gamma^r_{\theta\theta}(-\partial_r \Gamma^r_{\theta\theta})+2r-\mathbb{Q}r^3\big]\nonumber\\
&+g_{rr}r^3\frac{f(\mathbb{Q})}{f'(\mathbb{Q})}-2\big[(\Gamma^r_{\theta\theta})^2g_{rr}+2\Gamma^r_{\theta\theta}g_{rr}r+r^2\big]\frac{f''(\mathbb{Q})}{f'(\mathbb{Q})}(\partial_r\mathbb{Q})-2\big[(\Gamma^r_{\theta\theta})^2g_{rr}-r^2\big]\frac{f^{(3)}(\mathbb{Q})}{f''(\mathbb{Q})}(\partial_r\mathbb{Q})\Bigg\}.
\label{eq:EoMConnectionReduced}
\end{align}
\end{widetext}
%
%
%
Once again, equations (\ref{eq:ConstraintSolvedrthetatheta}) and (\ref{eq:EoMConnectionReduced}) correspond to those obtained in \cite{DAmbrosio2022} with the extra contributions coming from the coupling to matter. In particular, the EoM for the connection (\ref{eq:EoMConnectionReduced}) that we have obtained here by means of the Bianchi identities, is obtained in \cite{DAmbrosio2022} starting from (\ref{eq:EoMConnectionCovariant}), being the latter a considerably more convoluted task. 
The fact that the result presented here matches the one shown there serves both as a consistency check and as a motivation for introducing a much simpler and more elegant method for calculating such EoM.

As two further remarks, note that none of these equations depends on $\Chr{t}{\theta}{\theta}$. Consequently, the dynamics of the system are completely independent of this connection component. The quantity $\Chr{t}{\theta}{\theta}$ is therefore arbitrary and may be chosen, if desired, to take any constant value compatible with the geometrical postulates and the symmetries imposed on the system. Finally, it is important to note that the dependence on $\mathcal{T}_{tt}$ and $\mathcal{T}_{rr}$ in (\ref{eq:QSolvedForGammarrr}), (\ref{eq:ConstraintSolvedrthetatheta}), and (\ref{eq:EoMConnectionReduced}) only appears as a consequence of substituting $\partial_r g_{tt}$ and $\partial_r g_{rr}$ using their respective EoMs. Before this step—which we perform only for convenience and for the sake of comparison with \cite{DAmbrosio2022}—none of these equations depends on the energy-momentum tensor, since the connection does not couple to the matter fields.

In the vacuum limit, $\mathcal{T}_{\mu\nu}=0$ and equations (\ref{eq:EoMMetricgttReduced}), (\ref{eq:EoMMetricgrrReduced}), (\ref{eq:ConstraintSolvedrthetatheta}) and (\ref{eq:EoMConnectionReduced}) constitute a closed system of four equations for four unknowns: the three dynamical degrees of freedom $g_{tt}$, $g_{rr}$, $\Q$, and $\Chr{r}{\theta}{\theta}$. Concerning $\Chr{r}{\theta}{\theta}$, since it cannot be eliminated from the rest of the EoMs using (\ref{eq:ConstraintSolvedrthetatheta}), we solve also for it using this constraint equation to close the system.  However, once we consider a certain energy-momentum distribution described by some non-zero $\mathcal{T}_{\mu\nu}$, we have a larger number of variables to solve the system for. In the case of a NS, as will be detailed in the following sections, these new unknowns are the pressure of the star, $p(r)$, and its (energy) density, $\rho(r)$. Therefore, two new equations are necessary to solve the system unambiguously: a conservation equation for the energy-momentum tensor,
\begin{equation}
\mathring{\nabla}_\nu \mathcal{T}^{\mu\nu}=0;
\label{eq:ConservationEMomTensor}
\end{equation}
and a certain EoS:
\begin{equation}
f(p,\rho)=0.
\label{eq:EoSGeneric}
\end{equation}
The former corresponds to the same conservation law as in GR because the matter sector is assumed to be minimally coupled to the metric only, with no additional coupling to the independent connection or to the non-metricity scalar. The possible choices we can make for the latter and their physical motivation are discussed in detail in Sec.~\ref{sec:ChooseEoS}.

\subsection{Perfect fluid approximation and reparametrization of the metric tensor}
\label{sec:SphSols_PerfectFluid}

Concerning $\mathcal{T}_{\mu\nu}$, we assume for simplicity that our energy-momentum distribution is well modeled as a perfect fluid, i.e., transport processes such as heat conduction, viscosity or shear stress are negligible. This implies that perfect fluids are cold, ---there is no temperature dependence for $p$ and $\rho$--- and there are no viscosities; both scenarios happening in a NS after a reasonable amount of time has elapsed following the core collapse or merging of the source(s). We refer to Sec. \ref{sec:ChooseEoS} for further details.  The expression for $\mathcal{T}_{\mu\nu}$ that we consider from now on is then \cite{misner2017gravitation,Lin_2021}:
\begin{equation}
\mathcal{T}_{\mu \nu} = (\rho + p)u_{\mu}u_{\nu} + pg_{\mu\nu},
\label{eq:EnergyMomentumTensorPerfectFluid}
\end{equation}
where $u_{\mu}$ stands for the 4-velocity of the fluid elements and the pressure $p=p(r)$ and density $\rho=\rho(r)$ of the NS are functions of the radial distance to its center, $r$. Since we are considering a NS described as a self-gravitating perfect fluid, it has no motion and the only non-zero component of the 4-velocity is $u_{t}$. This, together with the normalization condition $u^{\mu}u_{\mu}=-1$ \cite{misner2017gravitation, schutz2009first}, is sufficient to determine (\ref{eq:EnergyMomentumTensorPerfectFluid}), given $g_{\mu\nu}$.\\

As far as the metric is concerned, one could treat $g_{tt}(r)$ and $g_{rr}(r)$ in (\ref{eq:ReducedMetric}) as the metric variables for which to solve the system and follow the approach in \cite{DAmbrosio2022}. However, it is common in the literature (e.g., in \cite{schutz2009first,misner2017gravitation, Lin_2021}) to introduce two functions $\xi(r)$ and $\zeta(r)$, so that the line element of a static, spherically symmetric spacetime is written as
\begin{equation}
ds^2=-e^{\xi (r)}dt^2+e^{\zeta (r)}dr^2+r^2(d\theta^2+\sin^2{\theta}\, d\phi^2).
\label{eq:SphericalMetricXiZeta}
\end{equation}
This reparametrization is acceptable provided that $g_{tt}<0$ and $g_{rr}>0$ at all points of the manifold, as would be expected for a NS. If one were to extend this analysis to generic (interior) BH solutions, note that these conditions are satisfied in the interior of stars, but they are in general not for BHs, for which it is necessary to revise this coordinate system and replace it with a more convenient alternative, such as the Kruskal-Szekeres coordinates, or use the generic notation with $g_{tt}$ and $g_{rr}$ as done in \cite{DAmbrosio2022}. 

Furthermore, for the sake of a later identification of $g_{rr}$ with the mass of the NS, we can further introduce a (not-yet-mass) $m(r)$ function so that:
\begin{equation}
	\zeta (r)\coloneqq \log\bigg(1-\frac{2m(r)}{r}\bigg)^{-1}.
	\label{eq:ZetaFunction}
\end{equation}
Under these assumptions, using (\ref{eq:EnergyMomentumTensorPerfectFluid}), (\ref{eq:SphericalMetricXiZeta}) and (\ref{eq:ZetaFunction}), the energy-momentum tensor $\mathcal{T}_{\mu\nu}$ reads:
\begin{equation}
\mathcal{T}_{\mu \nu} =\left(
\begin{array}{cccc}
 \rho (r)e^{\xi (r)} & 0 & 0 & 0 \\
 0 & p(r)e^{\zeta (r)} & 0 & 0 \\
 0 & 0 & p(r)r^2 & 0 \\
 0 & 0 & 0 & p(r)r^2 \sin ^2\theta \\
\end{array}
\right),
\end{equation}
and the conservation equation (\ref{eq:ConservationEMomTensor}) takes the form
\begin{equation}
\frac{d\xi}{dr}=-\frac{2}{p+\rho}\frac{dp}{dr}.
\label{eq:FinalEoMConservation}
\end{equation}
Note that this holds for any choice of connection components and $f(\mathbb{Q})$, since the covariant derivative is computed with respect to the Levi-Civita connection and it only depends on the matter fields and metric components. For clarifications in this regard, we refer to the proof of Bianchi identities in \cite{heisenberg2023reviewfqgravity}.

With all the above, the EoM for the metric (\ref{eq:EoMMetricgrrReduced}) and (\ref{eq:EoMMetricgttReduced}) turn into differential equations for the dynamical variables $m(r)$ and $\xi(r)$, where the latter can be used together with equation (\ref{eq:FinalEoMConservation}) to obtain an EoM for the pressure $p(r)$ instead and ``decouple'' the system from $\xi(r)$.  The connection EoM for $\Q$ and the constraint equation for $\Gamma^r_{\theta\theta}$ follow directly from (\ref{eq:EoMConnectionReduced}) and (\ref{eq:ConstraintSolvedrthetatheta}). These four equations take the form now:
\begin{widetext}
\begin{align}
\frac{dm}{dr}=&\frac{r^2}{2f'(\mathbb{Q})}\Bigg\{-\frac{1}{2}\big[f(\mathbb{Q})-f'(\mathbb{Q})\mathbb{Q}\big]+\frac{(\Gamma^r_{\theta\theta}+r)^2-2 m(2\Gamma^r_{\theta\theta}+r)}{r^2\Gamma^r_{\theta\theta}}(\partial_r\mathbb{Q})f''(\mathbb{Q})+\kappa\rho\Bigg\},
\label{eq:FinalEoMMass}\\[0.5em]
\frac{dp}{dr}=&-\frac{(p+\rho)}{2f'(\mathbb{Q})(r-2m)}\Bigg\{\frac{r^3 f(\mathbb{Q})+ f'(\mathbb{Q})\big(4m-r^3\mathbb{Q}\big)}{2r}-\frac{(\Gamma^r_{\theta\theta})^2+2rm-r^2}{\Gamma^r_{\theta\theta}}(\partial_r\mathbb{Q})f''(\mathbb{Q})+r^2\kappa p\Bigg\},
\label{eq:FinalEoMPressure}\\[0.5em]
\partial^2_r\mathbb{Q}=&\frac{(\partial_r\mathbb{Q})}{2\big[(\Gamma^r_{\theta\theta})^2+2rm-r^2\big]}\Bigg\{4(r-m)+4\Gamma^r_{\theta\theta}(-\partial_r\Gamma^r_{\theta\theta})-\mathbb{Q}r^3+r^3\frac{f(\mathbb{Q})}{f'(\mathbb{Q})}-2\big[(\Gamma^r_{\theta\theta}+r)^2-2rm\big]\frac{f''(\mathbb{Q})}{f'(\mathbb{Q})}(\partial_r\mathbb{Q})\nonumber\\
&-2\big[(\Gamma^r_{\theta\theta})^2+2rm-r^2\big]\frac{f^{(3)}(\mathbb{Q})}{f''(\mathbb{Q})}(\partial_r\mathbb{Q})\Bigg\}-\frac{r^2\kappa(\partial_r\mathbb{Q})}{2 f'(\mathbb{Q})(r-2m)}\Bigg\{p+\frac{(\Gamma^r_{\theta\theta})^2-2rm+r^2}{(\Gamma^r_{\theta\theta})^2+2rm-r^2}\rho\Bigg\},
\label{eq:FinalEoMQ}\\[0.5em]
\partial_r\Gamma^r_{\theta\theta}=&-\frac{\Gamma^r_{\theta\theta} \big\{f(\mathbb{Q})r^2(\Gamma^r_{\theta\theta}+r)+\big[2(r-2m)-(2 \Gamma^r_{\theta\theta}+r)(\mathbb{Q}r^2-2)\big]f'(\mathbb{Q})\big\}}{2 f'(\mathbb{Q}) \big[(\Gamma^r_{\theta\theta})^2+2rm-r^2\big]}\nonumber\\
&-\frac{r^2\kappa\Gamma^r_{\theta\theta}}{2 f'(\mathbb{Q})(r-2m)}\Bigg\{\rho+\frac{(\Gamma^r_{\theta\theta}+r)^2-2 m(2\Gamma^r_{\theta\theta}+r)}{(\Gamma^r_{\theta\theta})^2+2rm-r^2}p\Bigg\}.
\label{eq:FinalEoMGamma}
\end{align}
\end{widetext}
%
%
%
%
The system of equations $\{$(\ref{eq:FinalEoMConservation}), (\ref{eq:FinalEoMMass}), (\ref{eq:FinalEoMPressure}), (\ref{eq:FinalEoMQ}), (\ref{eq:FinalEoMGamma})$\}$, supplemented with a suitable EoS, constitutes the closed system of 6 equations for 6 unknown functions of $r$, namely $\{m,p, \rho, \xi,\Q;\Chr{r}{\theta}{\theta}\}$, which we want to solve for a given ansatz of the $f(\Q)$ function, as shall be done in Sec. \ref{sec:STEGRlim}, and \ref{sec:beyondGR}. We will refer to them from now on as our \textit{equations of structure}.

For the sake of completeness, one can also write down equation (\ref{eq:QSolvedForGammarrr}), which, if necessary, would allow us to obtain the remaining independent connection component once the previous system is solved:
\begin{widetext}
\begin{align}
\Gamma^r_{rr}&=\frac{\Gamma^r_{\theta\theta}f(\mathbb{Q})r^2(\Gamma^r_{\theta\theta}+r)+\big\{2r^2-4m(\Gamma^r_{\theta\theta}+r)+\Gamma^r_{\theta\theta}\big[2(2r+\Gamma^r_{\theta\theta})-r^2\mathbb{Q}(r+2\Gamma^r_{\theta\theta})\big]\big\}f'(\mathbb{Q})}{2f'(\mathbb{Q})\Gamma^r_{\theta\theta}\big[(\Gamma^r_{\theta\theta})^2+2rm-r^2\big]}\nonumber\\
&+\frac{r^2\kappa}{2f'(\mathbb{Q})(r-2m)}\Bigg\{\rho+\frac{(\Gamma^r_{\theta\theta}+r)^2-2m(2\Gamma^r_{\theta\theta}+r)}{(\Gamma^r_{\theta\theta})^2+2rm-r^2}p\Bigg\}.
\label{eq:FinalEoMGammaRRR}
\end{align}
\end{widetext}
\subsection{Choice of an EoS for neutron star solutions}\label{sec:ChooseEoS}
A NS is a type of star with typical masses of $M_{\rm NS}\sim 1-2$ $\mathrm{M}_{\odot}$, densities of the order of $\rho_{\rm NS}\sim 10^{14-15}$ $\mathrm{g/cm}^3$ and typical radii of $R_{\rm NS}\sim 10$ $\rm km$. Such stars do not burn nuclear fuel and are supported against gravity by the pressure of degenerate neutrons and nucleon–nucleon strong–interaction forces \cite{misner2017gravitation}. From this description it follows that its temperature is ideally fixed at zero and the nuclear composition can in principle be described entirely by its density. Therefore, a suitable EoS aspiring to describe a NS must be of the form $p(\rho)$, as anticipated at the beginning of Sec. \ref{sec:SphSols_PerfectFluid}. For the sake of simplicity, hereafter and when necessary we adopt in this work a polytropic EoS: 
\begin{equation}
p(r)=k\rho(r)^{\gamma}.
\label{eq:PolytropicEoS}
\end{equation}
This type of equations can be derived for any fluid under adiabatic conditions and depending on the value that the so-called adiabatic index $\gamma$ takes, it serves to describe a wide range of astrophysical objects. In particular, NS are well modeled as polytropes with $\gamma=2-3$ \cite{misner2017gravitation}.

This choice of EoS represents thus a deliberately simplified scenario and has been made mainly for practical and comparative reasons. In realistic NS modeling one typically employs tabulated microphysical EoS or piecewise–polytropic fits for different regions of the NS that capture composition changes across density ranges. While we do not adopt those here, extending our framework to realistic or piecewise–polytropic EoS is a natural next step left for future work. In particular, we highlight the interest of the approaches taken in \cite{O_Boyle_2020} and \cite{Read_2009}, to which we refer for more information in this regard.

\section{Understanding the STEGR limit}
\label{sec:STEGRlim}

Recent literature has revealed ambiguities regarding understanding which assumptions can be safely imposed when searching for beyond-GR solutions in $f(\Q)$ gravity. For instance, while setting the affine connection to zero is a valid gauge choice in STEGR, it generally eliminates additional degrees of freedom in the $f(\Q)$ scenario, suppressing the theory's distinctive features and potentially restricting the solution space to cases that fall within or closely resemble GR. Understanding the role of the connection in the STEGR limit is thus as a prerequisite for a consistent exploration of more general $f(\Q)$ models \cite{heisenberg2023reviewfqgravity}. This is particularly relevant when considering numerical approaches, where GR solutions often serve as a point of departure: knowing that the GR limit of a given system can be recovered numerically provides a robust consistency check and offers a concrete foundation from which to iteratively explore genuine beyond-GR behavior, as addressed in \ref{sec:beyondGR_numerics}. In this section, we comment on the general choices that inevitably retrieve the STEGR limit of the theory and establish a reference point for the subsequent numerical analysis of beyond-GR scenarios, where the connection may acquire a physically meaningful and dynamic role, as detailed in \ref{sec:SphSols_SymReduction}.\\

It is oftentimes adopted in the literature (e.g., \cite{Lin_2021,Alwan_2024}) the ansatz $\Chr{r}{\theta}{\theta}=-r$. This choice greatly simplifies our equations and from the constraint equation (\ref{eq:FinalEoMGamma}) we obtain the following expression for the non-metricity scalar:
\begin{equation}
	\mathbb{Q}(r)=-\frac{2\kappa m (p+\rho)}{f'(\mathbb{Q})(r-2m)}.
	\label{eq:NonMetricityScalarSphericalConnection}
\end{equation}
Furthermore, from making use of $\Chr{r}{\theta}{\theta}=-r$ together with (\ref{eq:NonMetricityScalarSphericalConnection}) in (\ref{eq:FinalEoMGammaRRR}), it follows that $\Chr{r}{r}{r}=0$.  Alternatively, we could have started by fixing the $\Chr{r}{r}{r}$ component to zero, solved (\ref{eq:FinalEoMGammaRRR}) for $\Q(r)$ and used these two results in (\ref{eq:FinalEoMGamma}) to end up getting $\Chr{r}{\theta}{\theta}=-r$ again. Substituting this into the expression obtained for $\Q(r)$ simplifies it back to (\ref{eq:NonMetricityScalarSphericalConnection}). In this sense, (\ref{eq:NonMetricityScalarSphericalConnection}) is given in general for the choice $\{\Chr{r}{\theta}{\theta}, \Chr{r}{r}{r}\}=\{-r, 0\}$. Alternative ways of writing this equation, adopted in different works are:
\begin{align}
	\Q&=-\frac{g_{rr}-1}{g_{tt}\,g_{rr}^2r}\,(g_{rr}\partial_r g_{tt}+g_{tt}\partial_r g_{rr}),\nonumber\\
          &=\frac{(e^{-\zeta}-1)(\zeta'+\xi')}{r};
	\label{eq:NonMetricityScalarSphericalConnectionAlternative}
\end{align}
being the former the result of computing the non-metricity scalar directly in terms of $g_{tt}$ and $g_{rr}$ using this connection; and the latter the expression for our reparametrization for the metric components (e.g., \cite{Lin_2021,nashed2024structuremaximummassstability,sharma2024physicalpropertiesmaximumcompactness,PhysRevD.105.024060,Junior_2024,DAmbrosio2022,Das_2024zsg,Pradhan_2024,Javed_2024}).

It can be trivially seen that for the vacuum case ($p=\rho=0$), which in turn means studying exterior NS solutions, $\Q=0$ and the connection field equation (\ref{eq:FinalEoMQ}) trivializes \cite{DAmbrosio2022}, losing then the associated additional degree of freedom and inevitably leading to GR results.
How de Sitter-Schwarzschild solutions arise from this scenario has been presented in \cite{DAmbrosio2022}, to which we refer for further details. In this regard, we can only stress that, for exterior solutions, adopting the spherical connection $\{\Chr{r}{\theta}{\theta}, \Chr{r}{r}{r}\}=\{-r, 0\}$ inherently leads to $\Q=0$ and consequently any attempt to use it to obtain beyond-GR solutions is doomed to failure \cite{DAmbrosio2022}. Therefore, when looking for such solutions, one must consider some sort of deviation from this behavior. This point will be further clarified in Sec. \ref{sec:beyondGR_constraints}.

When considering interior solutions, i.e., when $p$ and $\rho$ are different from zero; $\Q\neq0$, the EoM for $\Q$ is not identically satisfied and one obtains for the metric and connection EoMs, respectively:
\begin{widetext}
\begin{align}
\frac{dm}{dr}=&-\frac{r^2}{2f'(\mathbb{Q})}\bigg\{\frac{1}{2}\big[f(\mathbb{Q})-f'(\mathbb{Q})\mathbb{Q}\big]+\frac{2m}{r^2}(\partial_r\mathbb{Q})f''(\mathbb{Q})-\kappa\rho\bigg\},\label{eq:EoMmassSphericalConnection}\\
\frac{dp}{dr}=&-\frac{(p+\rho)}{2f'(\mathbb{Q})(r-2m)}\Bigg\{\frac{r^3 f(\mathbb{Q})+ f'(\mathbb{Q})\big(4m-r^3\mathbb{Q}\big)}{2r}+2m(\partial_r\mathbb{Q})f''(\mathbb{Q})+r^2\kappa p\Bigg\},\label{eq:EoMpressureSphericalConnection}\\
\partial^2_r\mathbb{Q}=&(\partial_r\mathbb{Q})\Bigg\{\frac{(r-m)\mathbb{Q}r^2}{4m^2}-\frac{1}{r}+\frac{r^2}{4m}\frac{f(\mathbb{Q})}{f'(\mathbb{Q})}+\frac{f''(\mathbb{Q})}{f'(\mathbb{Q})}(\partial_r\mathbb{Q})-\frac{f^{(3)}(\mathbb{Q})}{f''(\mathbb{Q})}(\partial_r\mathbb{Q})\Bigg\}+\frac{r^2\kappa(\partial_r\mathbb{Q})}{2 f'(\mathbb{Q})(r-2m)}\Bigg\{\rho+\bigg(\frac{r}{m}-1\bigg)p\Bigg\},
\label{eq:EoMConnectionSphericalConnection}
\end{align}
\end{widetext}
together with equation (\ref{eq:NonMetricityScalarSphericalConnection}) for the non-metricity scalar coming from (\ref{eq:FinalEoMGamma}) and (\ref{eq:FinalEoMGammaRRR}) as explained above.

The value of $\Q$ has not been substituted into the three equations above for the sake of clarity. We should note, however, that in this case, since one is artificially fixing all the independent components of the connection, this also fixes the form that $\Q$ must have and thus losing this additional degree of freedom. By fixing $\Q$ in this way, we might expect (\ref{eq:EoMConnectionSphericalConnection}) to lose all meaning and either impose some kind of restriction on the form that $f(\Q)$ can have in order to be satisfied identically, or most likely, to make the system of equations inconsistent, since we now have one less variable for the same number of equations and there is no guarantee that (\ref{eq:NonMetricityScalarSphericalConnection}) fulfills (\ref{eq:EoMConnectionSphericalConnection}). This would mean that such a choice of connection is directly incompatible with the kind of interior solutions we are looking for.
Again any attempt to fix only one of the two relevant independent components of the connection automatically fixes the other, making it impossible to find solutions with $\Chr{r}{\theta}{\theta}=-r$, $\Chr{r}{r}{r}\neq 0$ and vice versa: with $\Chr{r}{\theta}{\theta}\neq-r$, $\Chr{r}{r}{r}=0$. The strategies to be followed to avoid this outcome are again addressed in full detail in Sec. \ref{sec:beyondGR_constraints}.

Alternatively, and as briefly mentioned in Sec. \ref{sec:GeneralIntro_fQ}, the most direct possibility to recover the GR behavior of the theory without the need to specify any additional input is  to set $f(\Q)=\Q$, letting the independent connection components unspecified. By doing this in our EoMs for the metric (\ref{eq:FinalEoMPressure}) and (\ref{eq:FinalEoMMass}), we obtain:
\begin{align}
	\frac{dp}{dr}&=-\frac{(\rho+p)\big[m+(\kappa/2)pr^3\big]}{r(r-2m)},\label{eq:ToVEoMGr}\\
	\frac{dm}{dr}&=\frac{\kappa}{2}r^2\rho;\label{eq:MassEoMGr}
\end{align}
thus recovering the Tolman-Oppenheimer-Volkoff (TOV) and mass equations from GR without any extra assumptions for the connection. Naturally, it can also be trivially checked that by substituting $f(\Q)=\Q$ in (\ref{eq:EoMmassSphericalConnection}) and (\ref{eq:EoMpressureSphericalConnection}), these also lead to (\ref{eq:MassEoMGr}) and (\ref{eq:ToVEoMGr}) respectively. The two remaining equations to be considered to close the system are the one for the conservation of the energy-momentum tensor (\ref{eq:FinalEoMConservation}) and the EoS, for the variables $\{m,p,\rho,\xi\}$. 
As for the outer solutions, when setting $p=\rho=0$; (\ref{eq:ToVEoMGr}) is trivialized and (\ref{eq:FinalEoMConservation}) together with (\ref{eq:MassEoMGr}) lead to the Schwarzschild solution.

Concerning the role of the connection EoM in this scenario, and for later reference when commenting on how to depart from GR results numerically in Sec. \ref{sec:beyondGR_numerics}, we recall that equation (\ref{eq:EoMConnectionSphericalConnection}) originates precisely from the part of the Bianchi identities that is not automatically satisfied when deviating from GR, i.e., from the connection degree of freedom. Therefore, no analogous equation exists in the STEGR limit since the choice $f(\Q)=\Q$ trivializes it, and consequently, its associated EoM. Without delving into formal details---for which the reader is kindly referred once again to \cite{heisenberg2023reviewfqgravity}--- the derivation of (\ref{eq:EoMConnectionReduced}) demonstrates that the covariant form of the connection EoMs, $C_{\nu}=0$, arises from computing $\mathring{\nabla}_{\mu}\mathcal{M}\idx{^{\mu}_{\nu}}=0$, where $\mathcal{M}\idx{^{\mu}_{\nu}}$ is given by (\ref{eq:EoMMetricCovariant2}). However, in the STEGR limit, it follows from this equation that $\mathcal{M}\idx{^{\mu}_{\nu}}$ reduces to the Einstein tensor $\mathcal{G}\idx{^{\mu}_{\nu}}$, so that $\mathring{\nabla}_{\mu}\mathcal{G}\idx{^{\mu}_{\nu}}$ is identically zero, leading to the trivial tautology $0=0$ instead of a meaningful EoM.

Furthermore, due to the inherent gauge freedom in STEGR (see equation (\ref{eq:ConnectionGaugeFreedom})), the connection does not influence the physical dynamics, and therefore, in this limit, the EoMs for the metric remain invariant regardless of the connection choice. In practical terms, this implies that the problem can be fully addressed without specifying the connection at all, and hence the non-metricity scalar $\Q$, which remains physically irrelevant. Even if one opts to impose a specific ansatz for the connection in STEGR for convenience, and accordingly fixes the expression for $\Q$, any permissible connection choice (consistent with the imposed symmetries) would be related through a gauge transformation, rendering meaningless for the dynamics. As a token of remembrance, it is important to note that this gauge freedom is no longer present when transitioning to a more general $f(\Q)$ theory, provided that one does not want to artificially constrain the additional degree of freedom of the theory.

It is therefore theoretically not possible to find a well-defined (in the sense of being well-founded and unambiguous) GR limit for Eqs.~(\ref{eq:FinalEoMQ}) and (\ref{eq:FinalEoMGamma}). This leads to several complications, which will be discussed in detail in Sec.~\ref{sec:beyondGR_numerics}, and also results in discrepancies with some of the previous results in the literature, in particular those reported in Ref.~\cite{Lin_2021}.

\section{Beyond-GR neutron star solutions}
\label{sec:beyondGR}

The TOV and mass equations do not, in general, admit analytical solutions, and the same holds for the beyond-GR theories studied here. In this sense, we aim to obtain numerical solutions that can be confronted with observational data and, if necessary, discarded when they are inconsistent with current observations. An additional challenge in such modified scenarios is that the connection contributes non-trivially to the dynamics, so that we must not only solve the corresponding equations, but also specify initial conditions for $\Chr{r}{\theta}{\theta}$ and for $\Q$ ---whose EoM is of second order--- when integrating our system. Important comments on the viability of the models and on the issues that arise in the numerical integration are presented in Sec.~\ref{sec:beyondGR_numerics}. In the present section we study two different cosmology-motivated~\cite{heisenberg2023reviewfqgravity,Jim_nez_2020,ayuso2025insightsfqcosmologyrelevance} beyond-GR scenarios, characterized by different choices of $f(\Q)$, and analyze the behavior of the solutions at the integration boundaries, namely the stellar center and spatial infinity. This approach not only aids in assessing the regularity of the solutions and the internal consistency of the associated BVP, but also constrains the set of possible boundary conditions suitable for a numerical initial value problem (IVP) analysis. Moreover, we find interesting additional constraints on the solutions that eliminate the additional degrees of freedom associated with the connection. Finally, we review the various challenges encountered in the process of numerical integration, aiming to provide a practical walkthrough for future studies.

\subsection{EoMs for the cases $f(\Q)=\Q+\alpha\Q^2$ and $f(\Q)=\Q^{\beta}$}

The first family of theories we consider is the one given by $f(\Q)=\Q+\alpha\Q^2$, with $\alpha\in\mathbb{R}$ being a small parameter that characterizes the deviation from GR, i.e., from the case where $\alpha=0$. The equations of structure to be solved, particularized to our choice of $f(\Q)$ are: the metric field equations (\ref{eq:FinalEoMMass}) and (\ref{eq:FinalEoMPressure}), the connection EoM (\ref{eq:FinalEoMQ}), and the constraint equation (\ref{eq:FinalEoMGamma}), which read now:
\begin{widetext}
\begin{align}
\frac{dm}{dr}=&\frac{1}{2(1+2\alpha\mathbb{Q})}\Bigg\{\frac{\alpha}{2\Gamma^r_{\theta\theta}}\left\{r^2\mathbb{Q}^2\Gamma^r_{\theta\theta}+4(\partial_r\mathbb{Q})\big[(\Gamma^r_{\theta\theta}+r)^2-2m(2\Gamma^r_{\theta\theta}+r)\big]\right\}+\kappa r^2\rho\Bigg\},\label{eq:FinalEoMMassALPHA}\\[0.5em]
\frac{dp}{dr}=&\frac{(\rho+p)}{2(1+2\alpha\mathbb{Q})(r-2m)}\Bigg\{\frac{\alpha r^3\mathbb{Q}^2-4m(1+2\alpha\mathbb{Q})}{2r}+\frac{2\alpha\big[(\Gamma^r_{\theta\theta})^2+2rm-r^2\big](\partial_r\mathbb{Q})}{\Gamma^r_{\theta\theta}}-\kappa r^2p\Bigg\},
\label{eq:FinalEoMPressureALPHA}\\[0.5em]
\partial^2_r\mathbb{Q}=&-\frac{r^2\kappa(\partial_r\mathbb{Q})}{2(1+2\alpha\mathbb{Q})(r-2m)}\Bigg\{p+\frac{(\Gamma^r_{\theta\theta})^2-2rm+r^2}{(\Gamma^r_{\theta\theta})^2+2rm-r^2}\rho\Bigg\}+\frac{(\partial_r\mathbb{Q})}{1+2\alpha\mathbb{Q}}\Bigg\{\frac{2\big[r-m+\Gamma^r_{\theta\theta}(-\partial_r\Gamma^r_{\theta\theta})\big]}{(\Gamma^r_{\theta\theta})^2+2rm-r^2}\nonumber\\
&-\alpha\frac{\big[\mathbb{Q}^2r^3-8\mathbb{Q}\left(r-m+\Gamma^r_{\theta\theta}(-\partial_r\Gamma^r_{\theta\theta})\right)+4(\partial_r\mathbb{Q})\left((\Gamma^r_{\theta\theta}+r)^2-2rm\right)\big]}{2\big[(\Gamma^r_{\theta\theta})^2+2rm-r^2\big]}\Bigg\},
\label{eq:FinalEoMQALPHA}\\[0.5em]
\partial_r\Gamma^r_{\theta\theta}=&-\frac{\Gamma^r_{\theta\theta} \big\{4(r-m)+\Gamma^r_{\theta\theta}(4-r^2\mathbb{Q})+\alpha\mathbb{Q}\big[8(r-m+\Gamma^r_{\theta\theta})-r^2\mathbb{Q}\left(3\Gamma^r_{\theta\theta}+r\right)\big]\big\}}{2(1+2\alpha\mathbb{Q})\big[(\Gamma^r_{\theta\theta})^2+2rm-r^2\big]}\nonumber\\
&-\frac{r^2\kappa\Gamma^r_{\theta\theta}}{2(r-2m)(1+2\alpha\mathbb{Q})}\Bigg\{\rho+\frac{(\Gamma^r_{\theta\theta}+r)^2-2 m(2\Gamma^r_{\theta\theta}+r)}{(\Gamma^r_{\theta\theta})^2+2rm-r^2}p\Bigg\}.
\label{eq:FinalEoMGammaALPHA}
\end{align}
\end{widetext}
%
%

Moreover, it can be easily checked that (\ref{eq:MassEoMGr}) and (\ref{eq:ToVEoMGr}) are trivially recovered from (\ref{eq:FinalEoMMassALPHA}) and (\ref{eq:FinalEoMPressureALPHA}) respectively in the limit $\alpha = 0$.
It is also worth noting, especially now that the connection plays a fundamental role, that the EoM (\ref{eq:FinalEoMQALPHA}) is not only of second order but also that all terms on the right-hand side are proportional to $\partial_r \Q$. This implies that a solution where $\Q=\mathrm{const.}\neq 0$, while mathematically correct, would trivialize such equation, effectively guillotining the additional degree of freedom of the connection and, therefore, is not valid when aiming to find beyond-GR results. Likewise, we now also have equation (\ref{eq:FinalEoMGammaALPHA}), which, as previously noted, is not an EoM but rather a constraint equation for the $\Gamma^r_{\theta\theta}$ component that cannot be eliminated from the rest of the equations nor arbitrarily fixed, as discussed in Sec. \ref{sec:STEGRlim}. Given that in STEGR the connection is pure gauge and there is no analogous equation, it is both important and non-trivial to determine how to handle this equation and, in particular, to identify the appropriate boundary conditions not only for $\partial_r\Q$ but also for $\Gamma^r_{\theta\theta}$.

The second set of theories we consider is the one characterized by the function $f(\Q)=\Q^{\beta}$, $\beta\in\mathbb{R}$. On this occasion, the STEGR limit is obtained for $\beta=1$ and therefore we expect deviations from the GR behavior as much as $\beta$ differs from this value. The equations of structure including the matter read:
\begin{widetext}
\begin{align}
	\frac{dm}{dr}=&\frac{\beta-1}{4\beta}\Bigg\{\mathbb{Q}r^2+\frac{2\beta\big[(\Gamma^r_{\theta\theta}+r)^2-2m(2\Gamma^r_{\theta\theta}+r)\big]}{\mathbb{Q}\Gamma^r_{\theta\theta}}(\partial_r \mathbb{Q})\Bigg\}+\frac{\kappa r^2\mathbb{Q}^{1-\beta}}{2\beta}\rho,\\
	\frac{dp}{dr}=&\frac{(\rho+p)}{2\beta(r-2m)}\Bigg\{-\frac{2\beta m}{r}+(\beta-1)\Bigg[\frac{1}{2}\mathbb{Q}r^2+\frac{\beta\left((\Gamma^r_{\theta\theta})^2+2rm-r^2\right)}{\mathbb{Q}\Gamma^r_{\theta\theta}}(\partial_r \mathbb{Q})\Bigg]-\kappa r^2\mathbb{Q}^{1-\beta}p\Bigg\},
    \end{align}
\end{widetext}
\begin{widetext}
\begin{align}
	\partial^2_r\mathbb{Q}=&-\frac{r^2\kappa\mathbb{Q}^{1-\beta}(\partial_r\mathbb{Q})}{2\beta(r-2m)}\Bigg\{p+\frac{(\Gamma^r_{\theta\theta})^2-2rm+r^2}{(\Gamma^r_{\theta\theta})^2+2rm-r^2}\rho\Bigg\}-\frac{(\partial_r\mathbb{Q})}{\big[(\Gamma^r_{\theta\theta})^2+2rm-r^2\big]}\Bigg\{-2\big[r-m+\Gamma^r_{\theta\theta}(-\partial_r\Gamma^r_{\theta\theta})\big]\nonumber\\
	&+\frac{(\beta-1)}{2\beta}\mathbb{Q}r^3+\frac{\big[(\Gamma^r_{\theta\theta}+r)(r+\Gamma^r_{\theta\theta}(2\beta-3))-2rm\big]}{\mathbb{Q}}(\partial_r\mathbb{Q})\Bigg\},\\
    \partial_r\Gamma^r_{\theta\theta}=&-\frac{\Gamma^r_{\theta\theta} \big\{\mathbb{Q}r^2(\Gamma^r_{\theta\theta}+r)+\beta\big[4(r-m+\Gamma^r_{\theta\theta})-r^2\mathbb{Q}(2\Gamma^r_{\theta\theta}+r)\big]\big\}}{2\beta\big[(\Gamma^r_{\theta\theta})^2+2rm-r^2\big]}\nonumber\\
&-\frac{r^2\kappa\Gamma^r_{\theta\theta}\mathbb{Q}^{1-\beta}}{2\beta(r-2m)}\Bigg\{\rho+\frac{(\Gamma^r_{\theta\theta}+r)^2-2 m(2\Gamma^r_{\theta\theta}+r)}{(\Gamma^r_{\theta\theta})^2+2rm-r^2}p\Bigg\}.
\end{align}
\end{widetext}
As anticipated, the metric EoMs indeed recover Eqs.~(\ref{eq:ToVEoMGr}) and (\ref{eq:MassEoMGr}) in the limit $\beta = 1$. For the first set of theories considered, the parameter $\alpha$ enters only as an overall multiplicative factor in $f(\Q)$ and in its derivatives, so it merely rescales the coefficients of the equations without affecting their radial dependence. In contrast, $\beta$ appears as an exponent of the non-metricity scalar and therefore, when taking derivatives of $f(\Q)$, it generally modifies the order in $r$ of the different terms

\subsection{Constraints from regularity and boundary conditions}
\label{sec:beyondGR_constraints}

\begin{table*}[hbtp]
 \centering
 \caption{A summarized view of the initial conditions necessary to solve the BVP for the equations of structure of a NS in the STEGR limit ($f(\Q)=\Q$), together with the derived boundary conditions and Taylor expansion of their solution (leading-order and next-to-leading-order terms, up to $\mathcal{O}(r^3)$.}
   \renewcommand{\arraystretch}{1.5}
    \vspace{5pt}
  \begin{tabular}{|>{\centering\arraybackslash}S{m{4cm}}||>{\centering\arraybackslash}S{m{9cm}}|}
	\hline
   To be determined & $m^{(0)}$, $p^{(0)}$, $\xi^{(0)}$, ($\rho^{(0)}$ using EoS) \\
   \hline
	Boundary conditions & $m^{(0)}=0$, $p^{(0)}=\mathrm{const.}$, $\xi^{(0)}=\mathrm{const.}$, $\xi(r \rightarrow \infty)=0$ \\
   \hline
   \makecell{ Expansions at $r\approx 0\;$ \\ (truncated) } & \makecell{$m\approx\dfrac{\kappa \rho^{(0)}r^{3}}{6}$ $\textcolor{white}{X^{X^{X^{\big{|}}}}}$\\ $p\approx p^{(0)}-\dfrac{\kappa}{12}\left(\rho^{(0)}+p^{(0)}\right)\left(\rho^{(0)}+3p^{(0)}\right)r^2$ \\$\textcolor{white}{X_{X{X_{\big{|}}}}}$ $ \xi\approx \xi^{(0)}+\dfrac{\kappa}{6}\left(\rho^{(0)}+3p^{(0)}\right)r^2$}\\
   \hline
  \end{tabular}
  \label{Tab:GRInitialConditions}
  \vspace{5pt}
\end{table*}
In order to construct viable NS solutions and determine or constrain the space of possible initial conditions for the BVP, it is convenient to establish well-motivated physical assumptions that ensure the regularity of the solutions and the internal consistency of the boundary conditions with the equations of motion. A natural approach involves applying appropriate boundary conditions such as imposing regularity at the stellar center and ensuring asymptotic flatness at spatial infinity. We reason only on the three following assumptions:
\begin{enumerate}
\item The metric tensor and all other quantities must be regular and well-behaved at the center of the star, i.e., for $r=0$.
\item The solution must be asymptotically flat, i.e., the metric should be Minkowski-like for $r\rightarrow \infty$.
\item Both the quantities we are calculating and the corresponding differential equations admit a Maclaurin expansion in a neighborhood of $r=0$, and a Laurent expansion in powers of $1/r$ about $r\rightarrow \infty$.
\end{enumerate}
The first condition ensures a space geometry that is smooth at the origin. Any singularity at $r=0$ would imply that there is no local Lorentz there, yielding a physically unacceptable theory for NS. The second condition guarantees that infinitely far from the NS, gravity cancels out and one recovers the Special Relativity limit of the theory: a Minkowski spacetime. These conditions are thus physically-motivated and crucial to keep in mind if one wants to produce physically-relevant results. The third condition, while restricting the functional form of our solution to a specific ansatz, simplifies the study enormously without violating its analiticity and provides a straightforward path towards constraining the boundary conditions necessary to integrate the system numerically.

\subsubsection*{Interior solutions}
We consider the following perturbative analysis inside the NS:
\begin{align}
	p(r)&=p^{(0)}+p^{(1)}r+p^{(2)}r^2+\mathcal{O}(r^3),\nonumber\\
	\rho(r)&=\rho^{(0)}+\rho^{(1)}r+\rho^{(2)}r^2+\mathcal{O}(r^3),\nonumber\\
	m(r)&=m^{(0)}+m^{(1)}r+m^{(2)}r^2+\mathcal{O}(r^3),\nonumber\\
	\xi(r)&=\xi^{(0)}+\xi^{(1)}r+\xi^{(2)}r^2+\mathcal{O}(r^3), \nonumber \\
	\Chr{r}{\theta}{\theta}(r)&=\Chr{r}{\theta}{\theta}^{(0)}+\Chr{r}{\theta}{\theta}^{(1)}r+\Chr{r}{\theta}{\theta}^{(2)}r^2+\mathcal{O}(r^3),\nonumber \\
	\Q(r)&=\Q^{(0)}+\Q^{(1)}r+\Q^{(2)}r^2+\mathcal{O}(r^3);
	\label{eq:SystemExpansion_r}
\end{align}
where $A^{(0)}\equiv A(0)$, $A^{(n)}\equiv (d^nA/dr^n)_{r=0}$, for $A(r)$ being each one of the functions in (\ref{eq:SystemExpansion_r}). Next, we impose regularity of the equations of structure at $r=0$ by plugging the expansions above into our system of equations and determining, order by order, the conditions that must be met by the different coefficients in order to avoid divergences in the first place, and to fulfill the equation in second place.

For the STEGR case, the results of such study are well known \cite{misner2017gravitation} and we present them in TABLE \ref{Tab:GRInitialConditions} for later reference. We include the expansions (\ref{eq:SystemExpansion_r}) up to $\mathcal{O}(r^3)$, and a schematic overview of the required boundary conditions for the numerical integration, together with their constraints. In particular, we note that $m^{(0)}=0$ is required to ensure both a spacetime geometry smooth at the origin and regularity of the pressure equation (\ref{eq:ToVEoMGr}). The pressure $p^{(0)}\equiv p(0)$ at the center of the NS is an arbitrary constant of integration that can be fixed using available data from indirect observations or other well-tested models, as mentioned in \ref{sec:ChooseEoS}, and $\xi(0)$ is fixed by the junction condition when matching with the exterior Schwarzschild solution or in more general scenarios where the exterior solution is not known ---such as in beyond-GR cases---, by imposing an asymptotically flat metric, i.e., $\xi(r \rightarrow \infty)=0$.\\

For the two non-trivial choices of $f(\Q)$, the result of the analysis for the family of theories $f(\Q)=\Q+\alpha\Q^2$ is displayed in TABLE \ref{Tab:BeyondGRInitialConditionsALPHA}.
\begin{table*}[]
 \centering
 \caption{A summarized view of the initial conditions necessary to solve the BVP problem for the equations of structure of a NS in the beyond-GR case $f(\Q)=\Q+\alpha\Q^2$, together with the derived boundary conditions and Taylor expansion of their solution (leading-order and next-to-leading-order terms, up to $\mathcal{O}(r^3)$.}
  \renewcommand{\arraystretch}{1.5}
  \vspace{5pt}
  \begin{tabular}{|>{\centering\arraybackslash}S{m{4cm}}||>{\centering\arraybackslash}S{m{10.8cm}}|}
	\hline
   To be determined & $m^{(0)}$, $p^{(0)}$, $\xi^{(0)}$, $\Gamma^{r\;(0)}_{\theta\theta}$, $\mathbb{Q}^{(0)}$, $\mathbb{Q}^{(1)}$, ($\rho^{(0)}$ using EoS)\\
   \hline
	\makecell{$\textcolor{white}{a^{{{\big{|}}}}}$ Common boundary $\textcolor{white}{a^{{{\big{|}}}}}$\\ $\textcolor{white}{a_{{{\big{|}}}}}$ conditions $\textcolor{white}{a_{{{\big{|}}}}}$} & \makecell{$m^{(0)}=0$, $p^{(0)}=\mathrm{const.}$, $\xi^{(0)}=\mathrm{const.}$, $\xi(r \rightarrow \infty)=0$} \\
   \hline
   \makecell{ Common expansions\\ at $r\approx 0\;$ \\ (truncated) } & \makecell{$\textcolor{white}{X^{X^{X^{\big{|}}}}_{X_{X_{\big{|}}}}}$ $m\approx\dfrac{(\mathbb{Q}^{(0)})^2\alpha +2\rho^{(0)}\kappa}{6(4\alpha\mathbb{Q}^{(0)}+2)}r^3$ $\textcolor{white}{X^{X^{X^{\big{|}}}}_{X_{X_{\big{|}}}}}$\\$\textcolor{white}{X^{X^{X^{\big{|}}}}_{X_{X_{\big{|}}}}}$ $p\approx p^{(0)}+\dfrac{\left(\rho^{(0)}+p^{(0)}\right)\left[(\mathbb{Q}^{(0)})^2\alpha-\kappa\left(\rho^{(0)}+3p^{(0)}\right)\right]}{4(6\alpha\mathbb{Q}^{(0)}+3)}r^2$ $\textcolor{white}{X^{X^{X^{\big{|}}}}_{X_{X_{\big{|}}}}}$\\ $\textcolor{white}{X_{X_{X_{\big{|}}}}}$ $ \xi\approx \xi^{(0)}-\dfrac{(\mathbb{Q}^{(0)})^2\alpha-\kappa\left(\rho^{(0)}+3p^{(0)}\right)}{4(6\alpha\mathbb{Q}^{(0)}+3)}r^2$ $\textcolor{white}{X_{X_{X_{\big{|}}}}}$}\\
   \hline
  \end{tabular}
  \newline
\vspace*{0.2 cm}
\newline
    \begin{tabular}{|>{\centering\arraybackslash}S{m{4cm}}||>{\centering\arraybackslash}S{m{3.4cm}}|>{\centering\arraybackslash}S{m{3.7cm}}|>{\centering\arraybackslash}S{m{3.4cm}}|}
	\hline
	Different possibilities & Set 1 & Set 2 & Set 3 \\ 
	\hline
	Boundary conditions & \makecell{ $\textcolor{white}{|^{\big{|}}}$ $\Gamma^{r\;(0)}_{\theta\theta}=\mathrm{const.}$ $\textcolor{white}{|^{\big{|}}}$\\$\mathbb{Q}^{(0)}=\mathrm{const.}$\\$\textcolor{white}{|_{\big{|}}}$ $\mathbb{Q}^{(1)}=0$ $\textcolor{white}{|_{\big{|}}}$} & \makecell{$\textcolor{white}{|^{\big{|}}}$ $\Gamma^{r\;(0)}_{\theta\theta}=0$ $\textcolor{white}{|^{\big{|}}}$\\$\mathbb{Q}^{(0)}=\mathrm{const.}$\\$\textcolor{white}{|_{\big{|}}}$ $\mathbb{Q}^{(1)}=0$ $\textcolor{white}{|_{\big{|}}}$} & \makecell{$\textcolor{white}{|^{\big{|}}}$ $\Gamma^{r\;(0)}_{\theta\theta}=0$ $\textcolor{white}{|^{\big{|}}}$\\$\mathbb{Q}^{(0)}=0$\\$\textcolor{white}{|_{\big{|}}}$ $\mathbb{Q}^{(1)}=0$ $\textcolor{white}{|_{\big{|}}}$} \\
   \hline
   \makecell{$\textcolor{white}{|^{\big{|}}}$ Expansions at $\textcolor{white}{|^{\big{|}}}$\\ $\ \ r\approx 0\;$ (truncated) $\textcolor{white}{|_{\big{|}}}$} & \makecell{$\mathbb{Q}\approx \mathbb{Q}^{(0)}$\\$\Gamma^{r}_{\theta\theta}\approx \Gamma^{r\;(0)}_{\theta\theta}-2r$} & \makecell{$\mathbb{Q}\approx \mathbb{Q}^{(0)}$\\$\Gamma^{r}_{\theta\theta}\approx \Gamma^{r\;(2)}_{\theta\theta}r^2+\Gamma^{r\;(3)}_{\theta\theta}r^3$} & \makecell{$\mathbb{Q}\approx 0$\\$\Gamma^{r}_{\theta\theta}\approx -r+\Gamma^{r\;(2)}_{\theta\theta}r^2$} \\
   \hline
  \end{tabular}\textcolor{white}{***$*^*$}
   \label{Tab:BeyondGRInitialConditionsALPHA} 
\end{table*}
We observe that for the first case, the Maclaurin expansions for $m$, $p$ and $\xi$ are unambiguously determined and also consistently recover those presented in TABLE \ref{Tab:GRInitialConditions} for the STEGR case in the limit $\alpha =0$, serving as a consistency check. However, to maintain the regular behavior of the equations, the various choices for $\Chr{r}{\theta}{\theta}^{(0)}$ and $\Q^{(0)}$ give rise to three different sets of boundary conditions --- i.e., one can achieve this behavior by taking three different paths: The third set directly implies $\Q= 0$ and therefore recovers the STEGR scenario. For the other two sets, there exists a coupling between $m$, $p$ and $\xi$ and the non-metricity scalar. However, $\Q^{(1)}=0$ for both.\\

In fact, although only terms up to $\mathcal{O}(r^3)$ are shown in TABLE \ref{Tab:BeyondGRInitialConditionsALPHA}, our analysis shows that once $\Q^{(1)} = 0$ is fixed by regularity arguments, all subsequent derivatives also vanish. In other words, for sets 1 and 2, $\Q$ must be a constant. The immediate consequence of this result is that also here the connection EoM is trivialized everywhere inside the NS, therefore killing the associated degree of freedom. In this sense, while the equations for the metric and pressure appear to be coupled to the non-metricity scalar, this coupling lacks physical sense; it can be gauged away, rendering the analysis a convoluted reformulation of GR. This explains why different choices for the connection lead to the same solutions for the physical variables. The overarching conclusion of this analysis is that any attempt to derive a solution deviating from GR using the ansatz (\ref{eq:SystemExpansion_r}), i.e., a solution that admits a Taylor expansion around $r=0$, is inherently bound to fail and will inevitably converge to GR solutions.

For the beyond-GR scenario given by $f(\Q)=\Q^\beta$, the results of the analysis are gathered in TABLE \ref{Tab:BeyondGRInitialConditionsBETA} and we see that a similar situation occurs. Also here the Taylor expansions for $m$, $p$ and $\xi$ are determined unambiguously and recover the GR case in TABLE \ref{Tab:GRInitialConditions} for $\beta=1$, and different sets of initial conditions are obtained, depending on the values we take for $\Chr{r}{\theta}{\theta}^{(0)}$. In particular, the two sets obtained here are the same as sets 1 and 2 in TABLE \ref{Tab:BeyondGRInitialConditionsALPHA}. The conclusion for this case is thus completely analogous to that of the previous family of theories: we obtain that solutions that admit a Taylor series expansion around $r=0$ are only compatible with the situation where $\Q=\mathrm{const.}$ and therefore trivialize the EoM for the connection and can only lead to GR results.\\

\begin{table*}[]
 \centering
 \caption{A summarized view of the initial conditions necessary to solve the BVP problem for the equations of structure of a NS in the beyond-GR case $f(\Q)=\Q^{\beta}$, together with the derived boundary conditions and Taylor expansion of their solution (leading-order and next-to-leading-order terms, up to $\mathcal{O}(r^3)$.}
  \renewcommand{\arraystretch}{1.5}
  \vspace{5pt}
  \begin{tabular}{|>{\centering\arraybackslash}S{m{4cm}}||>{\centering\arraybackslash}S{m{10.8cm}}|}
	\hline
   To be determined & $m^{(0)}$, $p^{(0)}$, $\xi^{(0)}$, $\Gamma^{r\;(0)}_{\theta\theta}$, $\mathbb{Q}^{(0)}$, $\mathbb{Q}^{(1)}$, ($\rho^{(0)}$ using EoS)\\
   \hline
	\makecell{$\textcolor{white}{a^{{{\big{|}}}}}$ Common boundary $\textcolor{white}{a^{{{\big{|}}}}}$\\ $\textcolor{white}{a_{{{\big{|}}}}}$ conditions $\textcolor{white}{a_{{{\big{|}}}}}$} & \makecell{$m^{(0)}=0$, $p^{(0)}=\mathrm{const.}$, $\xi^{(0)}=\mathrm{const.}$, $\xi(r \rightarrow \infty)=0$} \\
   \hline
   \makecell{ Common expansions\\ at $r\approx 0\;$ \\ (truncated) } & \makecell{$\textcolor{white}{X^{X^{X^{\big{|}}}}_{X_{X_{\big{|}}}}}$ $m\approx\dfrac{\mathbb{Q}^{(0)}}{12\beta}\left[\beta-1+2\kappa\rho^{(0)}(\mathbb{Q}^{(0)})^{-\beta}\right]r^3$ $\textcolor{white}{X^{X^{X^{\big{|}}}}_{X_{X_{\big{|}}}}}$\\ $p\approx p^{(0)}+\dfrac{\mathbb{Q}^{(0)}}{12\beta}\left(\rho^{(0)}+p^{(0)}\right)\left[\beta-1-\kappa(\mathbb{Q}^{(0)})^{-\beta}\left(\rho^{(0)}+3p^{(0)}\right)\right]r^2$ \\ $\textcolor{white}{X^{X^{X^{\big{|}}}}_{X_{X_{\big{|}}}}}$ $ \xi\approx \xi^{(0)}-\dfrac{\mathbb{Q}^{(0)}}{6\beta}\left[\beta-1-\kappa(\mathbb{Q}^{(0)})^{-\beta}\left(\rho^{(0)}+3p^{(0)}\right)\right]r^2$ $\textcolor{white}{X^{X^{X^{\big{|}}}}_{X_{X_{\big{|}}}}}$}\\
   \hline
  \end{tabular}
  \newline
\vspace*{0.2 cm}
\newline
    \begin{tabular}{|>{\centering\arraybackslash}S{m{4cm}}||>{\centering\arraybackslash}S{m{5.32cm}}|>{\centering\arraybackslash}S{m{5.32cm}}|}
	\hline
	Different possibilities & Set 1 & Set \\ 
	\hline
	Boundary conditions & \makecell{ $\textcolor{white}{|^{\big{|}}}$ $\Gamma^{r\;(0)}_{\theta\theta}=\mathrm{const.}$ $\textcolor{white}{|^{\big{|}}}$\\$\mathbb{Q}^{(0)}=\mathrm{const.}$\\$\textcolor{white}{|_{\big{|}}}$ $\mathbb{Q}^{(1)}=0$ $\textcolor{white}{|_{\big{|}}}$} & \makecell{ $\textcolor{white}{|^{\big{|}}}$ $\Gamma^{r\;(0)}_{\theta\theta}=0$ $\textcolor{white}{|^{\big{|}}}$\\$\mathbb{Q}^{(0)}=\mathrm{const.}$\\$\textcolor{white}{|_{\big{|}}}$ $\mathbb{Q}^{(1)}=0$ $\textcolor{white}{|_{\big{|}}}$} \\
   \hline
   \makecell{$\textcolor{white}{|^{\big{|}}}$ Expansions at $\textcolor{white}{|^{\big{|}}}$\\ $\ \ r\approx 0\;$ (truncated) $\textcolor{white}{|_{\big{|}}}$} & \makecell{$\mathbb{Q}\approx \mathbb{Q}^{(0)}$ \\$\Gamma^{r}_{\theta\theta}\approx \Gamma^{r\;(0)}_{\theta\theta}-2r$} & \makecell{$\mathbb{Q}\approx \mathbb{Q}^{(0)}$ \\$\Gamma^{r}_{\theta\theta}\approx \Gamma^{r\;(2)}_{\theta\theta}r^2+\Gamma^{r\;(3)}_{\theta\theta}r^3$} \\
   \hline
  \end{tabular}\textcolor{white}{***$*^*$}
   \label{Tab:BeyondGRInitialConditionsBETA} 
\end{table*}
\subsubsection*{Exterior solutions}
For the outside of the NS, we explore the possibility of performing a series expansion in powers of $1/r$ around $r=\infty$, as one can do in GR to retrieve the Schwarzschild solution. In this way we intend to determine whether we are able to recover modified gravity effects to some order of $1/r$ where these become significant. This expansion is formulated as follows:
\begin{align}
	m(r)&=m^{(0)}+m^{(1)}r^{-1}+m^{(2)}r^{-2}+\mathcal{O}(r^{-3}),\nonumber\\
	\xi(r)&=\xi^{(0)}+\xi^{(1)}r^{-1}+\xi^{(2)}r^{-2}+\mathcal{O}(r^{-3}), \nonumber \\
	\Chr{r}{\theta}{\theta}(r)&=\Chr{r}{\theta}{\theta}^{(0)}+\Chr{r}{\theta}{\theta}^{(1)}r^{-1}+\Chr{r}{\theta}		{\theta}^{(2)}r^{-2}+\mathcal{O}(r^{-3}),\nonumber \\
	\Q(r)&=\Q^{(0)}+\Q^{(1)}r^{-1}+\Q^{(2)}r^{-2}+\mathcal{O}(r^{-3});
	\label{eq:SystemExpansion_OneOverR}
\end{align}
where $\xi^{(0)}=0$ and $\Q^{(0)}=0$ are required in order to ensure an asymptotic Minkowski spacetime.\\

Following the same steps outlined for the previous cases, we find also here that by applying this method, the only solution compatible with the asymptotic conditions and the functional dependence on $r$ in (\ref{eq:SystemExpansion_OneOverR}) for both families of beyond-GR theories is the trivial case where $\Q(r)=0$, corresponding to Schwarzschild's solution. This result poses a critical distinction from the STEGR scenario, where it is possible to recover the full exterior solution by means of such expansion.

\subsubsection*{General discussion}
In general, these results, while enforcing a further constraint on the shape of beyond-GR NS solutions within our setup, do not necessarily imply that such solutions do not exist. Instead, the analysis indicates that, among all the potential interior or exterior beyond-GR solutions for our system, none exhibit behavior as a power series in $r$ close to the center of the star or $1/r$ at far infinity, respectively. This approach assumes a specific dependence of the solutions on $r$ within the approximation range. Therefore, while one might be motivated to follow such an approach by the GR scenario, where this type of expansion is effective, in this more general case, the same reasoning does not hold. Consequently, adopting this kind of expansion will invariably lead to GR solutions, ruling out the possibility of observing genuinely beyond-GR effects in this regime.

As a counterexample and a reinforcing argument in favor of the possible existence of approximate beyond-GR solutions in the $f(\Q)=\Q+\alpha\Q^2$ case, we refer to the study conducted in \cite{DAmbrosio2022} on BH solutions. There, by performing a perturbative analysis in $\alpha$ rather than in $r$ the authors found an exterior solution that could be perfectly applicable to our NS scenario ---$\Chr{r}{\theta}{\theta}(r)$ is treated in their work as the dynamical variable, which translates into a third order differential EoM. The main advantage of such perturbative scheme is that it provides a picture of how much one must deviate from GR to appreciate the modified gravity effects and even obtain a scale at which this occurs. It is therefore a conceptually very transparent approach, even if not optimal for the computation of the ``fixed zones'' or the boundary conditions in the sense of what discussed above. 

%
%
%
%
%
%
%
%

In such exterior solution, the metric variables deviate from Schwarzschild by means of a connection hair with a radial dependence in $\log{r}/r$. Similarly, the non-metricity scalar shows a rather convoluted functional form, but not the metric nor $\Q$ diverge within the range of applicability, and asymptotically the metric takes the Minkowski form. Additionally, the non-metricity scalar has a well-defined limit at infinity: $\Q(r\rightarrow \infty)=0$. This behavior not only ensures an asymptotically flat spacetime, but also indicates that non-metricity vanishes at infinitely large distances. Since both curvature and torsion are zero by construction, these solutions guarantee a purely Minkowski spacetime devoid of gravity at infinity without the need for $\Q$ to trivialize anywhere outside the NS. The connection component $\Chr{r}{\theta}{\theta}(r)$ diverges at infinity, but so does it in STEGR, so this should not be regarded as a problem due to the previous arguments, as far as $\Q$ is well-behaved.

It is important to emphasize that this solution is likely not the only possible one and it is in principle possible to find other types of solutions by simply requiring that the energy-momentum tensor from the connection vanishes only at infinity, but not at every intermediate point from the surface of the sphere to there; i.e., that the EoM for the connection only trivializes at infinity. Furthermore, because of the non-trivial dependence on $r$ of the terms that deviate from GR to this order in perturbation theory, we can draw some conclusions as to why the expansion (\ref{eq:SystemExpansion_OneOverR}) fails to produce beyond-GR results: First of all, the physically meaningful spacetime quantities carrying dynamical degrees of freedom that we are considering are the metric and the non-metricity scalar. These are the geometrical objects that must be well-behaved at infinity and it follows from the results in \cite{DAmbrosio2022} that in order to do so, it is not necessary for $\Chr{r}{\theta}{\theta}$ to also be regular at infinity.\\

On the other hand and as anticipated above, the dominant contribution to the deviation from GR manifests itself in the form of a $\log{r}$ term, which does not admit a series expansion at infinity. In this sense, the complete solution could contain terms going like, e.g., $(\log{r})/r^n$ or $[r^n(A+B\log{r})]^{-1}$, and even $r^{-n}(C+D\log{r})$ for the connection ($\{A,B,C,D\}\in\mathbb{R}$); that we are overlooking in (\ref{eq:SystemExpansion_OneOverR}). As opposed to considering the now vast range of possibilities, looking for zeros in the energy-momentum tensor of the connection, far from being an easy job, might present a more productive alternative that we propose as future work.

Likewise for interior solutions, such beyond-GR behavior might emerge through more complex functional dependencies on $r$, such as $r\log{r}$ or $r^2/(1+r\log{r})$, which remain regular at the origin but do not support a Maclaurin expansion, in a similar manner to the behavior of exterior solutions.\\

\subsubsection*{A comment on extending the perturbation in $\alpha$}

For the sake of completeness and to provide a complementary approach to our study, we conclude the theoretical analysis by discussing the possibility of performing also here a perturbative analysis in $\alpha$ for the interior solutions in the first set of families. In doing so, the main conceptual differences that arise are, firstly, that the equations become considerably more complex in this case due to the appearance of $p(r)$ and $\rho(r)$. At the same time, this implies that one does not have an analytical solution for the zeroth-order approximation ---the STEGR limit--- as it happens with Schwarzschild's solution for the exterior case. One possible way to circumvent this issue would be to use the series expansions in TABLE \ref{Tab:GRInitialConditions} as such. 
On the other hand, the analysis in \cite{DAmbrosio2022} explores the deviation from GR by examining the deviation from the specific choice $\Chr{r}{\theta}{\theta}=-r$. As explained in Sec. \ref{sec:STEGRlim}, this choice effectively trivializes the connection EoM for exterior solutions but not for interior ones. In particular, for our choice of $f(\Q)$, this approach yields the result:
\begin{equation}
	\Q(r)=-\frac{1}{4\alpha}\left[1\pm\sqrt{1-\frac{16\kappa \alpha m\, (p+\rho)}{r-2m}}\right];
	\label{eq:NonMetricityScalarSphericalConnectionALPHA}
\end{equation}
where only the option with the minus sign has a well-defined limit when $\alpha\rightarrow 0$.

In principle, this does not pose any problem at the theoretical level since in STEGR the choice we make for the connection (and therefore for $\Q$) does not influence the dynamics, and there is no requirement to select it in any specific way. However, if we intend to analyze the behavior of (\ref{eq:FinalEoMQALPHA}) and (\ref{eq:FinalEoMGammaALPHA}) starting from the GR limit ---understood as the zeroth-order approximation of our solution--- we must select a scenario that simultaneously satisfies these two equations at $\mathcal{O}(\alpha^0)$ to serve as a valid starting point. 
Outside the NS, the choice $\Chr{r}{\theta}{\theta}^{(0)}=-r$ is a viable option that leads to a gauge where $\Q=0$. This choice significantly simplifies the equations and ensures a well-defined GR limit for both the EoM for the connection and the constraint equation for $\Chr{r}{\theta}{\theta}$. For interior solutions, since there is no apparent choice for $\Chr{r}{\theta}{\theta}^{(0)}(r)$ that implies $\Q=0$ or $\Q=\mathrm{const.}$, nor any alternative pair $\{\Chr{r}{\theta}{\theta}^{(0)}(r), \Q(r)\}$ that simultaneously satisfies (\ref{eq:FinalEoMQALPHA}) and (\ref{eq:FinalEoMGammaALPHA}), if one wishes to pursue this analysis, one must settle for simply finding an alternative choice for $\Chr{r}{\theta}{\theta}^{(0)}(r)$ that simplifies the analysis enough to enable a step-by-step resolution of the expansion. In our case, we tested the following \textit{Ansätze}:
\begin{align}
\Chr{r}{\theta}{\theta}^{(0)}(r)&=\{0, \, \Chr{r}{\theta}{\theta,c}, \,\Chr{r}{\theta}{\theta,c}\,r^4\log{r}, \,\pm r, \,\pm \sqrt{r\left(2m-r\right)},\nonumber\\
&\left(2m-r\right)\pm \sqrt{2m\left(2m-r\right)}\},
\end{align}
where $\Chr{r}{\theta}{\theta,c}$ denotes an arbitrary real constant. Nevertheless, the combined effect of the two aforementioned issues complicates the analytical resolution of the equations even within this approximation. Therefore, we do not believe that this approach offers a significant practical advantage in this particular case.



\subsection{Further remarks and guidelines on the numerical viability}
\label{sec:beyondGR_numerics}

In general terms, we solve the full beyond–GR system consisting of the metric equations~(\ref{eq:FinalEoMMass})–(\ref{eq:FinalEoMPressure}), the connection sector~(\ref{eq:FinalEoMQ})–(\ref{eq:FinalEoMGamma}), the conservation law~(\ref{eq:FinalEoMConservation}), and an EoS. Unlike the STEGR limit, where the connection is pure gauge and its EoM collapses to a tautology, here the non-metricity scalar $\Q$ satisfies a genuine second–order equation while $\Gamma^r_{\theta\theta}$ obeys a first–order constraint. 

We clarify the numerical pathologies encountered during the analysis that, in naïve integrations, tend to regress to GR-like configurations, and we present pragmatic ways to structure a consistent computation without suppressing the extra degree of freedom, serving as a reference for future studies.

\subsubsection*{GR attractor and boundary conditions}

A practical difficulty arises from the structure of (\ref{eq:FinalEoMQ}), where every term on its right–hand side is proportional to $\partial_r \Q$. Numerically, this makes the branch $\partial_r \Q\equiv 0$ an attractor: once the derivative is driven to zero by an initial guess or by round–off (for instance, due to coarse meshes, loose tolerances, or poor continuation), the update for $\Q$ stalls and the solver converges to $\Q=\mathrm{const.}$ (including $\Q=0$). The connection degree of freedom is thus effectively frozen, in line with the regularity results of Sec.~\ref{sec:beyondGR_constraints}, which show that Maclaurin–regular interiors force $\Q$ to be constant.

From a mathematical point of view, the static NS configurations considered here are naturally formulated as a BVP: regularity at $r=0$, matching conditions at the stellar surface, and asymptotic flatness as $r\to\infty$ constrain the solution both at the center and at large radii. In particular, asymptotic flatness requires $\Q(r\to\infty)=0$ (while the metric tends to Minkowski and the inertial connection may grow), so that $\Q(0)$ (and, in principle, also $\partial_r \Q(0)$ for a second–order equation) cannot be fixed from regularity alone and must be adjusted to satisfy the outer conditions. 

A straightforward numerical strategy is to cast the BVP into an IVP by shooting from the center: one treats the central values, such as $\{\Q(0),\,\partial_r\Q(0),\,\Gamma^r_{\theta\theta}(0)\}$, as free parameters, integrates the coupled four–equation system for $(m,p,\Q,\Gamma^r_{\theta\theta})$ outward (with $\xi$ determined from the conservation equation), and tunes them so that the surface and far–field conditions are met. In practice, however, the degeneracy of (\ref{eq:FinalEoMQ}) along the branch $\partial_r\Q\equiv 0$ can make simple shooting rather delicate: small variations in the central data, or the accumulation of numerical error, may push the solution towards the $\Q=\mathrm{const.}$ branch and drive the integration back to a GR–like regime.

For this reason, more robust schemes based on a global boundary–value formulation are likely preferable. While in our attempts we have restricted ourselves to a shooting implementation, future studies could instead discretize the radial domain, treat the central values $\{\Q(0),\,\partial_r\Q(0),\,\Gamma^r_{\theta\theta}(0)\}$ as unknowns to be solved for together with the bulk profiles, and impose regularity at $r=0$, the matching conditions at the stellar surface, and asymptotic flatness at a large but finite outer radius. In such a global BVP framework, parameter continuation in the deformation parameters that control departures from STEGR (for example, starting from a small but finite $\alpha=\alpha_\star\ll 1$ or $\beta=1+\epsilon$ with $\epsilon\neq 0$), together with suitably chosen initial guesses for $\Q(r)$ in the iterative solver, could help track branches with $\partial_r\Q\not\equiv 0$ and reduce the tendency of the method to converge to the constant–$\Q$ branch.

\subsubsection*{Interior anchoring, continuation, and exterior matching}

A second obstacle is the lack of an interior reference configuration for the connection sector. For the exterior, Sec.~\ref{sec:STEGRlim} provides a clear anchor: the choice $\Gamma^r_{\theta\theta}=-r$ implies $\Q=0$ via (\ref{eq:NonMetricityScalarSphericalConnection}). However, no analogous interior anchor exists. This means that in contrast to STEGR, there is no GR--like interior template for $\Gamma^r_{\theta\theta}$ or $\partial_r\Q$ that fixes their central behavior beyond regularity. This is relevant because (\ref{eq:FinalEoMQ}) is genuinely second order in $\Q$, whereas (\ref{eq:FinalEoMGamma}) is a first–order constraint for $\Gamma^r_{\theta\theta}$. 

Following the reasoning above, a practical way to remove this ambiguity is again to promote the central coefficients to unknowns determined by global boundary conditions. Concretely, one starts the integration not at $r=0$ but at $r=\varepsilon\ll 1$ and chooses the coefficient set that keeps $\partial_r\Q(\varepsilon)\neq 0$. The stellar radius $R$ is likewise unknown and is fixed by the free–boundary condition $p(R)=0$, enforced together with outer conditions that guarantee asymptotic flatness and vanishing nonmetricity. In this BVP setting, $\partial_r\Q(0)$ is thus not a tunable input but an outcome selected by simultaneously satisfying (\ref{eq:FinalEoMMass})–(\ref{eq:FinalEoMGamma}), the EoS, and the asymptotic (and observable) constraints. The same logic eliminates the need to impose boundary data for $\Gamma^r_{\theta\theta}$ by hand, since once its central value is included among the unknowns, the first–order constraint (\ref{eq:FinalEoMGamma}) propagates it throughout the domain, and the combination of central regularity and outer/observable conditions selects the admissible value. In addition, measured exterior observables can be folded into the outer boundary data (e.g.\ prescribing $(M,R)$ or the surface redshift $z_s$, or fixing the tidal Love number $\Lambda$), which translate into Dirichlet/Robin conditions for $(\xi,m,\Q)$ and help determine the missing central data.

A third source of difficulty is the ill–posed GR limit for the connection equations: in STEGR, $\mathcal{C}_\nu=0$ reduces to $\mathring{\nabla}_\mu \mathcal{G}\idx{^\mu_\nu}=0$ and (\ref{eq:FinalEoMGamma}) is absent, so there is no meaningful linearization point for (\ref{eq:FinalEoMQ})–(\ref{eq:FinalEoMGamma}). The aforementioned continuation should therefore start at a small nonzero deformation and the issue could potentially be fixed, for example, by proceeding with a pseudo–arclength continuation, rather than by dialing from zero; or by regularizing the connection EoM with a counterterm controlled by $\epsilon_{\rm reg}\to 0$ at convergence. 

Concerning exterior matching, we note that our early attempt to reproduce the exterior behavior suggested by the BH perturbative analysis by enforcing $\Q=0$ in the interior (and ignoring (\ref{eq:FinalEoMGamma}) there) while switching to the full system outside, leads to an apparent blow–up when integrating the mass equation. This is readily explained by the presence of logarithmic tails and indicates that such matching brings its own numerical pathology: logarithmic far–field terms (notably in the $\alpha$–family at ${\cal O}(\alpha^2)$) spoil naïve integration of the mass function when $m$ is used as the primary variable. Rather than shooting or solving for $m$ directly, this indicates that it might be preferable to work with $(\xi,\zeta)$ directly, or $(\xi,m)$ but subtract a known asymptotic template before integrating. If the latter were to  be done, one can write $m(r)=m_\infty+\delta m(r)$ and impose a Robin condition for $\delta m$ at a large $r_{\rm out}$ based on the template, or match at a finite $R_{\rm match}$ to the perturbative exterior for $(\xi,\Q)$, keeping in mind that only $\Gamma^r_{\theta\theta}$ may grow while the metric remains asymptotically flat and $\Q\to 0$.

Finally, if further issues appear near points where denominators such as $[(\Gamma^r_{\theta\theta})^2+2rm-r^2]$ or $\Gamma^r_{\theta\theta}$ become small, redefinitions involving dimensionless variables, and compactifying the radial domain so that the center behavior is regular by construction while remaining consistent with Sec.~\ref{sec:beyondGR_constraints} could potentially help.\\


\section{Conclusions and outlook}
\label{sec:conclusions}
\vspace{-0.6em}
This work was conceived to address a central issue in the study of NS within $f(\mathbb{Q})$ gravity: the consistent search for genuinely beyond-GR solutions that preserve the dynamical role of the affine connection.
While the field has seen growing interest —particularly in BH and cosmological contexts— many existing analyses of compact stars make simplifying assumptions, often introduced for technical convenience, which reduce or even remove the explicit role of the additional degrees of freedom inherent in $f(\mathbb{Q})$ gravity.  While some of these studies act as benchmark configurations within the $f(\mathbb{Q})$ framework and have played an important role in charting the space of solutions under such assumptions, in many cases, the gauge structure familiar from STEGR is implicitly carried over to more general models, so that the resulting configurations often remain very close to their GR counterparts. In this work, we have laid out a systematic framework —particularly oriented towards future, well-defined numerical studies— in which the connection is treated consistently as a physical, dynamical object throughout the construction, thereby complementing and extending these previous analyses towards setups where the affine degrees of freedom are fully taken into account. Building on earlier results in the context of BHs, we extended the formalism to spherically symmetric, non-vacuum configurations, employing a perfect fluid description with a polytropic EoS.

After briefly introducing the necessary elements of the formalism in Sec. \ref{sec:GeneralIntro}, we presented in \ref{sec:SphSols} the structure of the field equations for both the metric and the connection, adapting the results from \cite{DAmbrosio2022} to the NS case. Matter fields were included explicitly, and the system was closed using a versatile EoS relevant for astrophysical modeling. A key difference with respect to many earlier treatments of NS in the literature is that our setup retains a general connection consistent with the spacetime symmetries, which is not fixed a priori but allowed to be determined dynamically by the field equations. This distinguishes our work not only technically, but also conceptually: commonly used connection ansätze (often inherited from GR or STEGR) are very convenient in practice, but can restrict the dynamics in such a way that some of the genuinely extended behaviors permitted by $f(\mathbb{Q})$ gravity are not explored.

Sec. \ref{sec:STEGRlim} plays a brief but essential role. First, it clarifies the theoretical origin of a common issue: why many claimed ``beyond-GR” solutions actually lie within the GR regime due to hidden simplifications. More importantly, this section lays the groundwork for numerical exploration by identifying which limits must be recovered when looking for GR behavior and which equations serve as a starting point when seeking genuinely new physics. While concise, this section ensures that any future numerical implementation begins from a consistent and meaningful departure point. 

Thanks to this, it is in Sec. \ref{sec:beyondGR} where the core of the machinery to search for beyond-GR solutions is built. Focusing on two representative models —$f(\mathbb{Q}) = \mathbb{Q} + \alpha \mathbb{Q}^2$ and $f(\mathbb{Q}) = \mathbb{Q}^\beta$— we analyzed the structure of solutions near the stellar center and in the asymptotic regime motivated by the search of appropriate boundary conditions to solve the BVP. We found the key and constraining result that the beyond-GR branch cannot be captured by a regular expansion in pure power series in $r$ or $1/r$. While perhaps not a no-go theorem, this is a restrictive conclusion that places strong limitations on the admissible boundary behavior of any hypothetical beyond-GR configuration. It implies that genuinely novel effects cannot arise from standard perturbative methods routinely used in GR, but must instead originate from more intricate, possibly non-analytic features in the solution space —features that are systematically filtered out by the approximations most often employed. We also emphasized the distinct theoretical role of the connection's EoM. In both model scenarios studied, the behavior of this equation —particularly under regularity assumptions— acts as a diagnostic tool, helping determine whether a solution preserves the dynamical richness of the theory or collapses back into a GR-like regime. 

Lastly, in \ref{sec:beyondGR_numerics}, we reflected on the structural challenges that arise when attempting to solve the system numerically. We deliberately separate analytical groundwork from implementation to ensure that solver choices respect the constraints derived here, so that our analysis offers a clear set of criteria, both technical and conceptual, that should guide such future efforts.  First, the structure of Eq.~(\ref{eq:FinalEoMQ}), in which every term is proportional to $\partial_r \Q$, makes the branch $\partial_r\Q\equiv 0$ a numerical attractor, so that straightforward shooting implementations of the BVP tend to freeze the extra degree of freedom and drive the solution back to a GR–like regime. As discussed above, this suggests that genuinely beyond–GR branches might be more naturally approached within global boundary–value formulations, where the central data $\{\Q(0),\partial_r\Q(0),\Gamma^r_{\theta\theta}(0)\}$ are solved for together with the bulk profiles under regularity, matching, and asymptotic–flatness conditions, possibly aided by parameter continuation in the deformation parameters. Second, the absence of an interior GR–like anchor for the connection sector means that, unlike in STEGR, neither $\Gamma^r_{\theta\theta}$ nor $\partial_r\Q$ is fixed near the center beyond regularity requirements. In a global BVP setting, these quantities are instead selected by the combined constraints of the field equations and the interior and exterior boundary data, which can naturally incorporate observable quantities such as $(M,R)$, redshifts, or tidal deformabilities. Third, exterior matching must accommodate the logarithmic far-field structure, what favors either working with $(\xi,\zeta)$ instead of $m$, or solving for $m(r)=M_\infty+\delta m(r)$ with a Robin condition for $\delta m$ built from the perturbative template, or matching at a finite $R_{\rm match}$ to the known exterior for $(\xi,\Q)$—all while allowing $\Gamma^r_{\theta\theta}$ to grow inertially as the metric tends to Minkowski and $\Q\to0$.

If after implementing these strategies —or alternative ones proposed and refined by the community— within a genuinely well-defined BVP (with a regular center, asymptotically consistent outer data, and variable choices that neutralize the far-field pathology), the solutions still tend to fall back onto GR configurations, such an outcome would itself be informative about the robustness of the GR limit in these models, and we would welcome further suggestions and discussions on how best to probe this regime. This would indicate that, within static spherical symmetry, perfect-fluid matter without hypermomentum, and the most general flat, torsionless connection compatible with (\ref{eq:ReducedMetric}); the models $f(\Q)=\Q+\alpha\Q^2$ and $f(\Q)=\Q^\beta$ do not support Maclaurin/Laurent-regular beyond-GR neutron-star branches. Our message here is not a no-go claim, but a roadmap: the difficulties we exposed single out what a deeper, comprehensive numerical study must include (global BVPs, continuation, matched asymptotics, and observable-informed outer data) to faithfully probe the extra connection degree of freedom without oversimplifying it away. Should such a study still find no robust beyond-GR branch, the negative result would point either to a genuine obstruction under the stated assumptions (i.e., the connection’s extra mode decouples or becomes stealthy on static, isotropic fluid backgrounds) or to the need to relax them —rotation, anisotropy, time dependence, non-trivial matter couplings/hypermomentum, or less restrictive metric/connection ansätze— where non-analytic radial structure or different asymptotics might activate new physics.

To conclude, this work serves both as a theoretical clarification and as a strategic roadmap. It highlights that in $f(\mathbb{Q})$ gravity, obtaining truly novel behavior is not simply a matter of selecting a more complex function $f$, but of preserving the dynamical role of the affine structure and respecting the geometric subtleties of the theory. Many of the simplifying strategies commonly employed in the literature —often for good technical reasons— can nonetheless suppress precisely the features that distinguish these models from GR. In this work we have highlighted which assumptions tend to have this effect and provided criteria for avoiding them when the goal is to preserve the degrees of freedom introduced by the connection and to ensure that numerical schemes remain sensitive to genuinely beyond-GR dynamics. In this sense, we hope that this study complements existing analyses by contributing to a more systematic and transparent approach in the field, and helps set the stage for future investigations capable of genuinely new gravitational physics.

\vspace{-0.8em}
\section{Acknowledgements}
\vspace{-0.6em}
We thank Laur Järv and Tomi Koivisto for their useful comments and discussions. We are particularly grateful to Daniela D. Doneva and Stoytcho S. Yazadjiev for illuminating discussions, for the insightful input on the numerical aspects of this work, and for their exhaustive comments on the manuscript prior to submission.


\bibliographystyle{apsrev4-2} 
\bibliography{biblio}   

\begin{thebibliography}{72}%
\makeatletter
\providecommand \@ifxundefined [1]{%
 \@ifx{#1\undefined}
}%
\providecommand \@ifnum [1]{%
 \ifnum #1\expandafter \@firstoftwo
 \else \expandafter \@secondoftwo
 \fi
}%
\providecommand \@ifx [1]{%
 \ifx #1\expandafter \@firstoftwo
 \else \expandafter \@secondoftwo
 \fi
}%
\providecommand \natexlab [1]{#1}%
\providecommand \enquote  [1]{``#1''}%
\providecommand \bibnamefont  [1]{#1}%
\providecommand \bibfnamefont [1]{#1}%
\providecommand \citenamefont [1]{#1}%
\providecommand \href@noop [0]{\@secondoftwo}%
\providecommand \href [0]{\begingroup \@sanitize@url \@href}%
\providecommand \@href[1]{\@@startlink{#1}\@@href}%
\providecommand \@@href[1]{\endgroup#1\@@endlink}%
\providecommand \@sanitize@url [0]{\catcode `\\12\catcode `\$12\catcode `\&12\catcode `\#12\catcode `\^12\catcode `\_12\catcode `\%12\relax}%
\providecommand \@@startlink[1]{}%
\providecommand \@@endlink[0]{}%
\providecommand \url  [0]{\begingroup\@sanitize@url \@url }%
\providecommand \@url [1]{\endgroup\@href {#1}{\urlprefix }}%
\providecommand \urlprefix  [0]{URL }%
\providecommand \Eprint [0]{\href }%
\providecommand \doibase [0]{https://doi.org/}%
\providecommand \selectlanguage [0]{\@gobble}%
\providecommand \bibinfo  [0]{\@secondoftwo}%
\providecommand \bibfield  [0]{\@secondoftwo}%
\providecommand \translation [1]{[#1]}%
\providecommand \BibitemOpen [0]{}%
\providecommand \bibitemStop [0]{}%
\providecommand \bibitemNoStop [0]{.\EOS\space}%
\providecommand \EOS [0]{\spacefactor3000\relax}%
\providecommand \BibitemShut  [1]{\csname bibitem#1\endcsname}%
\let\auto@bib@innerbib\@empty
\bibitem [{\citenamefont {Heisenberg}(2019)}]{Heisenberg_2019}%
  \BibitemOpen
  \bibfield  {author} {\bibinfo {author} {\bibfnamefont {L.}~\bibnamefont {Heisenberg}},\ }\href {https://doi.org/10.1016/j.physrep.2018.11.006} {\bibfield  {journal} {\bibinfo  {journal} {Physics Reports}\ }\textbf {\bibinfo {volume} {796}},\ \bibinfo {pages} {1} (\bibinfo {year} {2019})}\BibitemShut {NoStop}%
\bibitem [{\citenamefont {Jimenez}\ \emph {et~al.}(2019)\citenamefont {Jimenez}, \citenamefont {Heisenberg},\ and\ \citenamefont {Koivisto}}]{GeomTrinity}%
  \BibitemOpen
  \bibfield  {author} {\bibinfo {author} {\bibfnamefont {J.~B.}\ \bibnamefont {Jimenez}}, \bibinfo {author} {\bibfnamefont {L.}~\bibnamefont {Heisenberg}},\ and\ \bibinfo {author} {\bibfnamefont {T.~S.}\ \bibnamefont {Koivisto}},\ }\href {https://doi.org/10.48550/ARXIV.1903.06830} {\bibinfo {title} {The geometrical trinity of gravity}} (\bibinfo {year} {2019})\BibitemShut {NoStop}%
\bibitem [{\citenamefont {Beltr{\'a}n~Jim{\'e}nez}\ \emph {et~al.}(2020{\natexlab{a}})\citenamefont {Beltr{\'a}n~Jim{\'e}nez}, \citenamefont {Heisenberg}, \citenamefont {Iosifidis}, \citenamefont {Jim{\'e}nez-Cano},\ and\ \citenamefont {Koivisto}}]{BeltranJimenez:2019odq}%
  \BibitemOpen
  \bibfield  {author} {\bibinfo {author} {\bibfnamefont {J.}~\bibnamefont {Beltr{\'a}n~Jim{\'e}nez}}, \bibinfo {author} {\bibfnamefont {L.}~\bibnamefont {Heisenberg}}, \bibinfo {author} {\bibfnamefont {D.}~\bibnamefont {Iosifidis}}, \bibinfo {author} {\bibfnamefont {A.}~\bibnamefont {Jim{\'e}nez-Cano}},\ and\ \bibinfo {author} {\bibfnamefont {T.~S.}\ \bibnamefont {Koivisto}},\ }\href {https://doi.org/10.1016/j.physletb.2020.135422} {\bibfield  {journal} {\bibinfo  {journal} {Phys. Lett. B}\ }\textbf {\bibinfo {volume} {805}},\ \bibinfo {pages} {135422} (\bibinfo {year} {2020}{\natexlab{a}})},\ \Eprint {https://arxiv.org/abs/1909.09045} {arXiv:1909.09045 [gr-qc]} \BibitemShut {NoStop}%
\bibitem [{\citenamefont {Jim{\'{e}}nez}\ \emph {et~al.}(2018{\natexlab{a}})\citenamefont {Jim{\'{e}}nez}, \citenamefont {Heisenberg},\ and\ \citenamefont {Koivisto}}]{Jim_nez_2018}%
  \BibitemOpen
  \bibfield  {author} {\bibinfo {author} {\bibfnamefont {J.~B.}\ \bibnamefont {Jim{\'{e}}nez}}, \bibinfo {author} {\bibfnamefont {L.}~\bibnamefont {Heisenberg}},\ and\ \bibinfo {author} {\bibfnamefont {T.}~\bibnamefont {Koivisto}},\ }\bibfield  {journal} {\bibinfo  {journal} {Physical Review D}\ }\textbf {\bibinfo {volume} {98}},\ \href {https://doi.org/10.1103/physrevd.98.044048} {10.1103/physrevd.98.044048} (\bibinfo {year} {2018}{\natexlab{a}})\BibitemShut {NoStop}%
\bibitem [{\citenamefont {Jim{\'{e}}nez}\ \emph {et~al.}(2018{\natexlab{b}})\citenamefont {Jim{\'{e}}nez}, \citenamefont {Heisenberg},\ and\ \citenamefont {Koivisto}}]{Jim_nez_2018_2}%
  \BibitemOpen
  \bibfield  {author} {\bibinfo {author} {\bibfnamefont {J.~B.}\ \bibnamefont {Jim{\'{e}}nez}}, \bibinfo {author} {\bibfnamefont {L.}~\bibnamefont {Heisenberg}},\ and\ \bibinfo {author} {\bibfnamefont {T.~S.}\ \bibnamefont {Koivisto}},\ }\href {https://doi.org/10.1088/1475-7516/2018/08/039} {\bibfield  {journal} {\bibinfo  {journal} {Journal of Cosmology and Astroparticle Physics}\ }\textbf {\bibinfo {volume} {2018}}\bibinfo  {number} { (08)},\ \bibinfo {pages} {039}}\BibitemShut {NoStop}%
\bibitem [{\citenamefont {Heisenberg}(2023)}]{heisenberg2023reviewfqgravity}%
  \BibitemOpen
\bibfield  {number} {  }\bibfield  {author} {\bibinfo {author} {\bibfnamefont {L.}~\bibnamefont {Heisenberg}},\ }\href {https://arxiv.org/abs/2309.15958} {\bibinfo {title} {Review on $f(q)$ gravity}} (\bibinfo {year} {2023}),\ \Eprint {https://arxiv.org/abs/2309.15958} {arXiv:2309.15958 [gr-qc]} \BibitemShut {NoStop}%
\bibitem [{\citenamefont {Heisenberg}\ \emph {et~al.}(2024)\citenamefont {Heisenberg}, \citenamefont {Hohmann},\ and\ \citenamefont {Kuhn}}]{Heisenberg:2023wgk}%
  \BibitemOpen
  \bibfield  {author} {\bibinfo {author} {\bibfnamefont {L.}~\bibnamefont {Heisenberg}}, \bibinfo {author} {\bibfnamefont {M.}~\bibnamefont {Hohmann}},\ and\ \bibinfo {author} {\bibfnamefont {S.}~\bibnamefont {Kuhn}},\ }\href {https://doi.org/10.1088/1475-7516/2024/03/063} {\bibfield  {journal} {\bibinfo  {journal} {JCAP}\ }\textbf {\bibinfo {volume} {03}},\ \bibinfo {pages} {063}},\ \Eprint {https://arxiv.org/abs/2311.05495} {arXiv:2311.05495 [gr-qc]} \BibitemShut {NoStop}%
\bibitem [{\citenamefont {D'Ambrosio}\ \emph {et~al.}(2020{\natexlab{a}})\citenamefont {D'Ambrosio}, \citenamefont {Garg}, \citenamefont {Heisenberg},\ and\ \citenamefont {Zentarra}}]{DAmbrosio:2020nqu}%
  \BibitemOpen
  \bibfield  {author} {\bibinfo {author} {\bibfnamefont {F.}~\bibnamefont {D'Ambrosio}}, \bibinfo {author} {\bibfnamefont {M.}~\bibnamefont {Garg}}, \bibinfo {author} {\bibfnamefont {L.}~\bibnamefont {Heisenberg}},\ and\ \bibinfo {author} {\bibfnamefont {S.}~\bibnamefont {Zentarra}},\ }\href@noop {} {\  (\bibinfo {year} {2020}{\natexlab{a}})},\ \Eprint {https://arxiv.org/abs/2007.03261} {arXiv:2007.03261 [gr-qc]} \BibitemShut {NoStop}%
\bibitem [{\citenamefont {D'Ambrosio}\ \emph {et~al.}(2023)\citenamefont {D'Ambrosio}, \citenamefont {Heisenberg},\ and\ \citenamefont {Zentarra}}]{DAmbrosio:2023asf}%
  \BibitemOpen
  \bibfield  {author} {\bibinfo {author} {\bibfnamefont {F.}~\bibnamefont {D'Ambrosio}}, \bibinfo {author} {\bibfnamefont {L.}~\bibnamefont {Heisenberg}},\ and\ \bibinfo {author} {\bibfnamefont {S.}~\bibnamefont {Zentarra}},\ }\href {https://doi.org/10.1002/prop.202300185} {\bibfield  {journal} {\bibinfo  {journal} {Fortsch. Phys.}\ }\textbf {\bibinfo {volume} {71}},\ \bibinfo {pages} {2300185} (\bibinfo {year} {2023})},\ \Eprint {https://arxiv.org/abs/2308.02250} {arXiv:2308.02250 [gr-qc]} \BibitemShut {NoStop}%
\bibitem [{\citenamefont {Heisenberg}(2025)}]{heisenberg2025countingdegreesfreedommethod}%
  \BibitemOpen
  \bibfield  {author} {\bibinfo {author} {\bibfnamefont {L.}~\bibnamefont {Heisenberg}},\ }\href {https://arxiv.org/abs/2509.18192} {\bibinfo {title} {Counting degrees of freedom: A method applicable from scalars to f(q) gravity and beyond}} (\bibinfo {year} {2025}),\ \Eprint {https://arxiv.org/abs/2509.18192} {arXiv:2509.18192 [math-ph]} \BibitemShut {NoStop}%
\bibitem [{\citenamefont {D'Ambrosio}\ \emph {et~al.}(2021)\citenamefont {D'Ambrosio}, \citenamefont {Heisenberg},\ and\ \citenamefont {Kuhn}}]{DAmbrosio2021}%
  \BibitemOpen
  \bibfield  {author} {\bibinfo {author} {\bibfnamefont {F.}~\bibnamefont {D'Ambrosio}}, \bibinfo {author} {\bibfnamefont {L.}~\bibnamefont {Heisenberg}},\ and\ \bibinfo {author} {\bibfnamefont {S.}~\bibnamefont {Kuhn}},\ }\href {https://doi.org/10.1088/1361-6382/ac3f99} {\bibfield  {journal} {\bibinfo  {journal} {Classical and Quantum Gravity}\ }\textbf {\bibinfo {volume} {39}},\ \bibinfo {pages} {025013} (\bibinfo {year} {2021})}\BibitemShut {NoStop}%
\bibitem [{\citenamefont {D'Ambrosio}\ \emph {et~al.}(2022)\citenamefont {D'Ambrosio}, \citenamefont {Fell}, \citenamefont {Heisenberg},\ and\ \citenamefont {Kuhn}}]{DAmbrosio2022}%
  \BibitemOpen
  \bibfield  {author} {\bibinfo {author} {\bibfnamefont {F.}~\bibnamefont {D'Ambrosio}}, \bibinfo {author} {\bibfnamefont {S.~D.~B.}\ \bibnamefont {Fell}}, \bibinfo {author} {\bibfnamefont {L.}~\bibnamefont {Heisenberg}},\ and\ \bibinfo {author} {\bibfnamefont {S.}~\bibnamefont {Kuhn}},\ }\bibfield  {journal} {\bibinfo  {journal} {Physical Review D}\ }\textbf {\bibinfo {volume} {105}},\ \href {https://doi.org/10.1103/physrevd.105.024042} {10.1103/physrevd.105.024042} (\bibinfo {year} {2022})\BibitemShut {NoStop}%
\bibitem [{\citenamefont {Rünkla}\ and\ \citenamefont {Vilson}(2018)}]{R_nkla_2018}%
  \BibitemOpen
  \bibfield  {author} {\bibinfo {author} {\bibfnamefont {M.}~\bibnamefont {Rünkla}}\ and\ \bibinfo {author} {\bibfnamefont {O.}~\bibnamefont {Vilson}},\ }\bibfield  {journal} {\bibinfo  {journal} {Physical Review D}\ }\textbf {\bibinfo {volume} {98}},\ \href {https://doi.org/10.1103/physrevd.98.084034} {10.1103/physrevd.98.084034} (\bibinfo {year} {2018})\BibitemShut {NoStop}%
\bibitem [{\citenamefont {D'Ambrosio}\ \emph {et~al.}(2020{\natexlab{b}})\citenamefont {D'Ambrosio}, \citenamefont {Garg},\ and\ \citenamefont {Heisenberg}}]{DAmbrosio:2020nev}%
  \BibitemOpen
  \bibfield  {author} {\bibinfo {author} {\bibfnamefont {F.}~\bibnamefont {D'Ambrosio}}, \bibinfo {author} {\bibfnamefont {M.}~\bibnamefont {Garg}},\ and\ \bibinfo {author} {\bibfnamefont {L.}~\bibnamefont {Heisenberg}},\ }\href {https://doi.org/10.1016/j.physletb.2020.135970} {\bibfield  {journal} {\bibinfo  {journal} {Phys. Lett. B}\ }\textbf {\bibinfo {volume} {811}},\ \bibinfo {pages} {135970} (\bibinfo {year} {2020}{\natexlab{b}})},\ \Eprint {https://arxiv.org/abs/2004.00888} {arXiv:2004.00888 [gr-qc]} \BibitemShut {NoStop}%
\bibitem [{\citenamefont {Dimakis}\ \emph {et~al.}(2022)\citenamefont {Dimakis}, \citenamefont {Paliathanasis}, \citenamefont {Roumeliotis},\ and\ \citenamefont {Christodoulakis}}]{PhysRevD.106.043509}%
  \BibitemOpen
  \bibfield  {author} {\bibinfo {author} {\bibfnamefont {N.}~\bibnamefont {Dimakis}}, \bibinfo {author} {\bibfnamefont {A.}~\bibnamefont {Paliathanasis}}, \bibinfo {author} {\bibfnamefont {M.}~\bibnamefont {Roumeliotis}},\ and\ \bibinfo {author} {\bibfnamefont {T.}~\bibnamefont {Christodoulakis}},\ }\href {https://doi.org/10.1103/PhysRevD.106.043509} {\bibfield  {journal} {\bibinfo  {journal} {Phys. Rev. D}\ }\textbf {\bibinfo {volume} {106}},\ \bibinfo {pages} {043509} (\bibinfo {year} {2022})}\BibitemShut {NoStop}%
\bibitem [{\citenamefont {Chen}(2025)}]{Chen_2025}%
  \BibitemOpen
  \bibfield  {author} {\bibinfo {author} {\bibfnamefont {W.-X.}\ \bibnamefont {Chen}},\ }\href {https://doi.org/10.1139/cjp-2024-0214} {\bibfield  {journal} {\bibinfo  {journal} {Canadian Journal of Physics}\ }\textbf {\bibinfo {volume} {103}},\ \bibinfo {pages} {531–542} (\bibinfo {year} {2025})}\BibitemShut {NoStop}%
\bibitem [{\citenamefont {Nashed}\ and\ \citenamefont {Saridakis}(2025)}]{nashed20253dimensionalchargedblackholes}%
  \BibitemOpen
  \bibfield  {author} {\bibinfo {author} {\bibfnamefont {G.~G.~L.}\ \bibnamefont {Nashed}}\ and\ \bibinfo {author} {\bibfnamefont {E.~N.}\ \bibnamefont {Saridakis}},\ }\href {https://arxiv.org/abs/2506.10046} {\bibinfo {title} {3-dimensional charged black holes in $f({Q})$ gravity}} (\bibinfo {year} {2025}),\ \Eprint {https://arxiv.org/abs/2506.10046} {arXiv:2506.10046 [gr-qc]} \BibitemShut {NoStop}%
\bibitem [{\citenamefont {Junior}\ and\ \citenamefont {Rodrigues}(2023)}]{Junior_2023}%
  \BibitemOpen
  \bibfield  {author} {\bibinfo {author} {\bibfnamefont {J.~T. S.~S.}\ \bibnamefont {Junior}}\ and\ \bibinfo {author} {\bibfnamefont {M.~E.}\ \bibnamefont {Rodrigues}},\ }\bibfield  {journal} {\bibinfo  {journal} {The European Physical Journal C}\ }\textbf {\bibinfo {volume} {83}},\ \href {https://doi.org/10.1140/epjc/s10052-023-11660-2} {10.1140/epjc/s10052-023-11660-2} (\bibinfo {year} {2023})\BibitemShut {NoStop}%
\bibitem [{\citenamefont {Junior}\ \emph {et~al.}(2024)\citenamefont {Junior}, \citenamefont {Lobo},\ and\ \citenamefont {Rodrigues}}]{Junior_2024}%
  \BibitemOpen
  \bibfield  {author} {\bibinfo {author} {\bibfnamefont {J.~T. S.~S.}\ \bibnamefont {Junior}}, \bibinfo {author} {\bibfnamefont {F.~S.~N.}\ \bibnamefont {Lobo}},\ and\ \bibinfo {author} {\bibfnamefont {M.~E.}\ \bibnamefont {Rodrigues}},\ }\bibfield  {journal} {\bibinfo  {journal} {The European Physical Journal C}\ }\textbf {\bibinfo {volume} {84}},\ \href {https://doi.org/10.1140/epjc/s10052-024-12696-8} {10.1140/epjc/s10052-024-12696-8} (\bibinfo {year} {2024})\BibitemShut {NoStop}%
\bibitem [{\citenamefont {Gogoi}\ \emph {et~al.}(2023)\citenamefont {Gogoi}, \citenamefont {Övgün},\ and\ \citenamefont {Koussour}}]{Gogoi_2023}%
  \BibitemOpen
  \bibfield  {author} {\bibinfo {author} {\bibfnamefont {D.~J.}\ \bibnamefont {Gogoi}}, \bibinfo {author} {\bibfnamefont {A.}~\bibnamefont {Övgün}},\ and\ \bibinfo {author} {\bibfnamefont {M.}~\bibnamefont {Koussour}},\ }\bibfield  {journal} {\bibinfo  {journal} {The European Physical Journal C}\ }\textbf {\bibinfo {volume} {83}},\ \href {https://doi.org/10.1140/epjc/s10052-023-11881-5} {10.1140/epjc/s10052-023-11881-5} (\bibinfo {year} {2023})\BibitemShut {NoStop}%
\bibitem [{\citenamefont {Nashed}(2024)}]{Nashed_2024}%
  \BibitemOpen
  \bibfield  {author} {\bibinfo {author} {\bibfnamefont {G.~G.~L.}\ \bibnamefont {Nashed}},\ }\bibfield  {journal} {\bibinfo  {journal} {Fortschritte der Physik}\ }\textbf {\bibinfo {volume} {72}},\ \href {https://doi.org/10.1002/prop.202400037} {10.1002/prop.202400037} (\bibinfo {year} {2024})\BibitemShut {NoStop}%
\bibitem [{\citenamefont {Nashed}(2025)}]{nashed2025specialndimensionalchargedantidesitter}%
  \BibitemOpen
  \bibfield  {author} {\bibinfo {author} {\bibfnamefont {G.~G.~L.}\ \bibnamefont {Nashed}},\ }\href {https://arxiv.org/abs/2312.14451} {\bibinfo {title} {Special $n$-dimensional charged anti-de-sitter black holes in $f(\mathbb{Q})$ gravitational theory}} (\bibinfo {year} {2025}),\ \Eprint {https://arxiv.org/abs/2312.14451} {arXiv:2312.14451 [gr-qc]} \BibitemShut {NoStop}%
\bibitem [{\citenamefont {Zhang}\ \emph {et~al.}(2025)\citenamefont {Zhang}, \citenamefont {Lan},\ and\ \citenamefont {Miao}}]{zhang2025commentblackholesfmathbbq}%
  \BibitemOpen
  \bibfield  {author} {\bibinfo {author} {\bibfnamefont {Z.-X.}\ \bibnamefont {Zhang}}, \bibinfo {author} {\bibfnamefont {C.}~\bibnamefont {Lan}},\ and\ \bibinfo {author} {\bibfnamefont {Y.-G.}\ \bibnamefont {Miao}},\ }\href {https://arxiv.org/abs/2508.12912} {\bibinfo {title} {Comment on "black holes in $f(\mathbb{Q})$ gravity"}} (\bibinfo {year} {2025}),\ \Eprint {https://arxiv.org/abs/2508.12912} {arXiv:2508.12912 [gr-qc]} \BibitemShut {NoStop}%
\bibitem [{\citenamefont {Nashed}\ and\ \citenamefont {Harko}(2024)}]{nashed2024structuremaximummassstability}%
  \BibitemOpen
  \bibfield  {author} {\bibinfo {author} {\bibfnamefont {G.~G.~L.}\ \bibnamefont {Nashed}}\ and\ \bibinfo {author} {\bibfnamefont {T.}~\bibnamefont {Harko}},\ }\href {https://arxiv.org/abs/2410.13968} {\bibinfo {title} {Structure, maximum mass, and stability of compact stars in f(q,t) gravity}} (\bibinfo {year} {2024}),\ \Eprint {https://arxiv.org/abs/2410.13968} {arXiv:2410.13968 [gr-qc]} \BibitemShut {NoStop}%
\bibitem [{\citenamefont {Sharma}\ \emph {et~al.}(2024)\citenamefont {Sharma}, \citenamefont {Ghosh},\ and\ \citenamefont {Paul}}]{sharma2024physicalpropertiesmaximumcompactness}%
  \BibitemOpen
  \bibfield  {author} {\bibinfo {author} {\bibfnamefont {R.}~\bibnamefont {Sharma}}, \bibinfo {author} {\bibfnamefont {A.}~\bibnamefont {Ghosh}},\ and\ \bibinfo {author} {\bibfnamefont {A.}~\bibnamefont {Paul}},\ }\href {https://arxiv.org/abs/2409.04487} {\bibinfo {title} {Physical properties and the maximum compactness bound of a class of compact stars in $f(q)$ gravity}} (\bibinfo {year} {2024}),\ \Eprint {https://arxiv.org/abs/2409.04487} {arXiv:2409.04487 [gr-qc]} \BibitemShut {NoStop}%
\bibitem [{\citenamefont {de~Araujo}\ and\ \citenamefont {Fortes}(2024)}]{dearaujo2024compactstarsfq}%
  \BibitemOpen
  \bibfield  {author} {\bibinfo {author} {\bibfnamefont {J.~C.~N.}\ \bibnamefont {de~Araujo}}\ and\ \bibinfo {author} {\bibfnamefont {H.~G.~M.}\ \bibnamefont {Fortes}},\ }\href {https://arxiv.org/abs/2407.08884} {\bibinfo {title} {Compact stars in $f(q) = q +\xi q^2$ gravity}} (\bibinfo {year} {2024}),\ \Eprint {https://arxiv.org/abs/2407.08884} {arXiv:2407.08884 [gr-qc]} \BibitemShut {NoStop}%
\bibitem [{\citenamefont {Maurya}\ \emph {et~al.}(2024)\citenamefont {Maurya}, \citenamefont {Newton~Singh}, \citenamefont {Mustafa}, \citenamefont {Govender}, \citenamefont {Errehymy},\ and\ \citenamefont {Aziz}}]{Maurya_2024}%
  \BibitemOpen
  \bibfield  {author} {\bibinfo {author} {\bibfnamefont {S.}~\bibnamefont {Maurya}}, \bibinfo {author} {\bibfnamefont {K.}~\bibnamefont {Newton~Singh}}, \bibinfo {author} {\bibfnamefont {G.}~\bibnamefont {Mustafa}}, \bibinfo {author} {\bibfnamefont {M.}~\bibnamefont {Govender}}, \bibinfo {author} {\bibfnamefont {A.}~\bibnamefont {Errehymy}},\ and\ \bibinfo {author} {\bibfnamefont {A.}~\bibnamefont {Aziz}},\ }\href {https://doi.org/10.1088/1475-7516/2024/09/048} {\bibfield  {journal} {\bibinfo  {journal} {Journal of Cosmology and Astroparticle Physics}\ }\textbf {\bibinfo {volume} {2024}}\bibinfo  {number} { (09)},\ \bibinfo {pages} {048}}\BibitemShut {NoStop}%
\bibitem [{\citenamefont {Maurya}\ \emph {et~al.}(2022{\natexlab{a}})\citenamefont {Maurya}, \citenamefont {Singh}, \citenamefont {Lohakare},\ and\ \citenamefont {Mishra}}]{Maurya_2022}%
  \BibitemOpen
\bibfield  {number} {  }\bibfield  {author} {\bibinfo {author} {\bibfnamefont {S.~K.}\ \bibnamefont {Maurya}}, \bibinfo {author} {\bibfnamefont {K.~N.}\ \bibnamefont {Singh}}, \bibinfo {author} {\bibfnamefont {S.~V.}\ \bibnamefont {Lohakare}},\ and\ \bibinfo {author} {\bibfnamefont {B.}~\bibnamefont {Mishra}},\ }\bibfield  {journal} {\bibinfo  {journal} {Fortschritte der Physik}\ }\textbf {\bibinfo {volume} {70}},\ \href {https://doi.org/10.1002/prop.202200061} {10.1002/prop.202200061} (\bibinfo {year} {2022}{\natexlab{a}})\BibitemShut {NoStop}%
\bibitem [{\citenamefont {Lin}\ and\ \citenamefont {Zhai}(2021)}]{Lin_2021}%
  \BibitemOpen
  \bibfield  {author} {\bibinfo {author} {\bibfnamefont {R.-H.}\ \bibnamefont {Lin}}\ and\ \bibinfo {author} {\bibfnamefont {X.-H.}\ \bibnamefont {Zhai}},\ }\bibfield  {journal} {\bibinfo  {journal} {Physical Review D}\ }\textbf {\bibinfo {volume} {103}},\ \href {https://doi.org/10.1103/physrevd.103.124001} {10.1103/physrevd.103.124001} (\bibinfo {year} {2021})\BibitemShut {NoStop}%
\bibitem [{\citenamefont {Das}\ and\ \citenamefont {Chattopadhyay}(2025)}]{Das_2024zsg}%
  \BibitemOpen
  \bibfield  {author} {\bibinfo {author} {\bibfnamefont {S.}~\bibnamefont {Das}}\ and\ \bibinfo {author} {\bibfnamefont {S.}~\bibnamefont {Chattopadhyay}},\ }\href {https://doi.org/10.1016/j.astropartphys.2024.103053} {\bibfield  {journal} {\bibinfo  {journal} {Astropart. Phys.}\ }\textbf {\bibinfo {volume} {165}},\ \bibinfo {pages} {103053} (\bibinfo {year} {2025})}\BibitemShut {NoStop}%
\bibitem [{\citenamefont {Alwan}\ \emph {et~al.}(2024)\citenamefont {Alwan}, \citenamefont {Inagaki}, \citenamefont {Mishra},\ and\ \citenamefont {Narawade}}]{Alwan_2024}%
  \BibitemOpen
  \bibfield  {author} {\bibinfo {author} {\bibfnamefont {M.~A.}\ \bibnamefont {Alwan}}, \bibinfo {author} {\bibfnamefont {T.}~\bibnamefont {Inagaki}}, \bibinfo {author} {\bibfnamefont {B.}~\bibnamefont {Mishra}},\ and\ \bibinfo {author} {\bibfnamefont {S.}~\bibnamefont {Narawade}},\ }\href {https://doi.org/10.1088/1475-7516/2024/09/011} {\bibfield  {journal} {\bibinfo  {journal} {Journal of Cosmology and Astroparticle Physics}\ }\textbf {\bibinfo {volume} {2024}}\bibinfo  {number} { (09)},\ \bibinfo {pages} {011}}\BibitemShut {NoStop}%
\bibitem [{\citenamefont {Maurya}\ \emph {et~al.}(2023{\natexlab{a}})\citenamefont {Maurya}, \citenamefont {Singh}, \citenamefont {Govender}, \citenamefont {Mustafa},\ and\ \citenamefont {Ray}}]{maurya2023effectgravitationaldecouplingconstraining}%
  \BibitemOpen
\bibfield  {number} {  }\bibfield  {author} {\bibinfo {author} {\bibfnamefont {S.~K.}\ \bibnamefont {Maurya}}, \bibinfo {author} {\bibfnamefont {K.~N.}\ \bibnamefont {Singh}}, \bibinfo {author} {\bibfnamefont {M.}~\bibnamefont {Govender}}, \bibinfo {author} {\bibfnamefont {G.}~\bibnamefont {Mustafa}},\ and\ \bibinfo {author} {\bibfnamefont {S.}~\bibnamefont {Ray}},\ }\href {https://arxiv.org/abs/2309.10130} {\bibinfo {title} {The effect of gravitational decoupling on constraining the mass and radius for the secondary component of gw190814 and other self-bound strange stars in f(q)-gravity theory}} (\bibinfo {year} {2023}{\natexlab{a}}),\ \Eprint {https://arxiv.org/abs/2309.10130} {arXiv:2309.10130 [gr-qc]} \BibitemShut {NoStop}%
\bibitem [{\citenamefont {Pradhan}\ and\ \citenamefont {Sahoo}(2024)}]{Pradhan_2024}%
  \BibitemOpen
  \bibfield  {author} {\bibinfo {author} {\bibfnamefont {S.}~\bibnamefont {Pradhan}}\ and\ \bibinfo {author} {\bibfnamefont {P.}~\bibnamefont {Sahoo}},\ }\href {https://doi.org/10.1016/j.nuclphysb.2024.116523} {\bibfield  {journal} {\bibinfo  {journal} {Nuclear Physics B}\ }\textbf {\bibinfo {volume} {1002}},\ \bibinfo {pages} {116523} (\bibinfo {year} {2024})}\BibitemShut {NoStop}%
\bibitem [{\citenamefont {Alwan}\ \emph {et~al.}(2025)\citenamefont {Alwan}, \citenamefont {Inagaki}, \citenamefont {Narawade},\ and\ \citenamefont {Mishra}}]{10.1093mnrasstaf1999}%
  \BibitemOpen
  \bibfield  {author} {\bibinfo {author} {\bibfnamefont {M.~A.}\ \bibnamefont {Alwan}}, \bibinfo {author} {\bibfnamefont {T.}~\bibnamefont {Inagaki}}, \bibinfo {author} {\bibfnamefont {S.~A.}\ \bibnamefont {Narawade}},\ and\ \bibinfo {author} {\bibfnamefont {B.}~\bibnamefont {Mishra}},\ }\href {https://doi.org/10.1093/mnras/staf1999} {\bibfield  {journal} {\bibinfo  {journal} {Monthly Notices of the Royal Astronomical Society}\ ,\ \bibinfo {pages} {staf1999}} (\bibinfo {year} {2025})},\ \Eprint {https://arxiv.org/abs/https://academic.oup.com/mnras/advance-article-pdf/doi/10.1093/mnras/staf1999/65284741/staf1999.pdf} {https://academic.oup.com/mnras/advance-article-pdf/doi/10.1093/mnras/staf1999/65284741/staf1999.pdf} \BibitemShut {NoStop}%
\bibitem [{\citenamefont {Sharif}\ and\ \citenamefont {Ajmal}(2025)}]{Sharif_2025}%
  \BibitemOpen
  \bibfield  {author} {\bibinfo {author} {\bibfnamefont {M.}~\bibnamefont {Sharif}}\ and\ \bibinfo {author} {\bibfnamefont {M.}~\bibnamefont {Ajmal}},\ }\bibfield  {journal} {\bibinfo  {journal} {Fortschritte der Physik}\ }\textbf {\bibinfo {volume} {73}},\ \href {https://doi.org/10.1002/prop.202400225} {10.1002/prop.202400225} (\bibinfo {year} {2025})\BibitemShut {NoStop}%
\bibitem [{\citenamefont {Ibrar}\ and\ \citenamefont {Sharif}(2025)}]{ibrar2025analyzinggravastarstructurefinchskea}%
  \BibitemOpen
  \bibfield  {author} {\bibinfo {author} {\bibfnamefont {I.}~\bibnamefont {Ibrar}}\ and\ \bibinfo {author} {\bibfnamefont {M.}~\bibnamefont {Sharif}},\ }\href {https://arxiv.org/abs/2502.09679} {\bibinfo {title} {Analyzing gravastar structure with the finch-skea metric in extended modified symmetric teleparallel gravity}} (\bibinfo {year} {2025}),\ \Eprint {https://arxiv.org/abs/2502.09679} {arXiv:2502.09679 [gr-qc]} \BibitemShut {NoStop}%
\bibitem [{\citenamefont {Mohanty}\ and\ \citenamefont {Sahoo}(2024)}]{Mohanty_2024}%
  \BibitemOpen
  \bibfield  {author} {\bibinfo {author} {\bibfnamefont {D.}~\bibnamefont {Mohanty}}\ and\ \bibinfo {author} {\bibfnamefont {P.~K.}\ \bibnamefont {Sahoo}},\ }\bibfield  {journal} {\bibinfo  {journal} {Fortschritte der Physik}\ }\textbf {\bibinfo {volume} {72}},\ \href {https://doi.org/10.1002/prop.202400082} {10.1002/prop.202400082} (\bibinfo {year} {2024})\BibitemShut {NoStop}%
\bibitem [{\citenamefont {Mohanty}\ \emph {et~al.}(2024)\citenamefont {Mohanty}, \citenamefont {Ghosh},\ and\ \citenamefont {Sahoo}}]{Mohanty_2024_2}%
  \BibitemOpen
  \bibfield  {author} {\bibinfo {author} {\bibfnamefont {D.}~\bibnamefont {Mohanty}}, \bibinfo {author} {\bibfnamefont {S.}~\bibnamefont {Ghosh}},\ and\ \bibinfo {author} {\bibfnamefont {P.}~\bibnamefont {Sahoo}},\ }\href {https://doi.org/10.1016/j.aop.2024.169636} {\bibfield  {journal} {\bibinfo  {journal} {Annals of Physics}\ }\textbf {\bibinfo {volume} {463}},\ \bibinfo {pages} {169636} (\bibinfo {year} {2024})}\BibitemShut {NoStop}%
\bibitem [{\citenamefont {Javed}\ \emph {et~al.}(2024)\citenamefont {Javed}, \citenamefont {Waseem}, \citenamefont {Mustafa}, \citenamefont {Tchier}, \citenamefont {Atamurotov}, \citenamefont {Ahmedov},\ and\ \citenamefont {Abdujabbarov}}]{Javed_2024}%
  \BibitemOpen
  \bibfield  {author} {\bibinfo {author} {\bibfnamefont {F.}~\bibnamefont {Javed}}, \bibinfo {author} {\bibfnamefont {A.}~\bibnamefont {Waseem}}, \bibinfo {author} {\bibfnamefont {G.}~\bibnamefont {Mustafa}}, \bibinfo {author} {\bibfnamefont {F.}~\bibnamefont {Tchier}}, \bibinfo {author} {\bibfnamefont {F.}~\bibnamefont {Atamurotov}}, \bibinfo {author} {\bibfnamefont {B.}~\bibnamefont {Ahmedov}},\ and\ \bibinfo {author} {\bibfnamefont {A.}~\bibnamefont {Abdujabbarov}},\ }\href {https://doi.org/10.1016/j.cjph.2024.04.022} {\bibfield  {journal} {\bibinfo  {journal} {Chinese Journal of Physics}\ }\textbf {\bibinfo {volume} {90}},\ \bibinfo {pages} {410–421} (\bibinfo {year} {2024})}\BibitemShut {NoStop}%
\bibitem [{\citenamefont {Pradhan}\ \emph {et~al.}(2023{\natexlab{a}})\citenamefont {Pradhan}, \citenamefont {Mohanty},\ and\ \citenamefont {Sahoo}}]{Pradhan_2023}%
  \BibitemOpen
  \bibfield  {author} {\bibinfo {author} {\bibfnamefont {S.}~\bibnamefont {Pradhan}}, \bibinfo {author} {\bibfnamefont {D.}~\bibnamefont {Mohanty}},\ and\ \bibinfo {author} {\bibfnamefont {P.}~\bibnamefont {Sahoo}},\ }\href {https://doi.org/10.1088/1674-1137/ace311} {\bibfield  {journal} {\bibinfo  {journal} {Chinese Physics C}\ }\textbf {\bibinfo {volume} {47}},\ \bibinfo {pages} {095104} (\bibinfo {year} {2023}{\natexlab{a}})}\BibitemShut {NoStop}%
\bibitem [{\citenamefont {Pradhan}\ \emph {et~al.}(2023{\natexlab{b}})\citenamefont {Pradhan}, \citenamefont {Mandal},\ and\ \citenamefont {Sahoo}}]{Pradhan_2023_2}%
  \BibitemOpen
  \bibfield  {author} {\bibinfo {author} {\bibfnamefont {S.}~\bibnamefont {Pradhan}}, \bibinfo {author} {\bibfnamefont {S.}~\bibnamefont {Mandal}},\ and\ \bibinfo {author} {\bibfnamefont {P.}~\bibnamefont {Sahoo}},\ }\href {https://doi.org/10.1088/1674-1137/acc1ce} {\bibfield  {journal} {\bibinfo  {journal} {Chinese Physics C}\ }\textbf {\bibinfo {volume} {47}},\ \bibinfo {pages} {055103} (\bibinfo {year} {2023}{\natexlab{b}})}\BibitemShut {NoStop}%
\bibitem [{\citenamefont {Bhattacharjee}\ and\ \citenamefont {Chattopadhyay}(2025)}]{bhattacharjee2025}%
  \BibitemOpen
  \bibfield  {author} {\bibinfo {author} {\bibfnamefont {D.}~\bibnamefont {Bhattacharjee}}\ and\ \bibinfo {author} {\bibfnamefont {P.~K.}\ \bibnamefont {Chattopadhyay}},\ }\href {https://arxiv.org/abs/2505.17583} {\bibinfo {title} {Exploring gravastar-like structures with strongly interacting quark matter shell in the framework of $f(q)$ gravity under conformal symmetry}} (\bibinfo {year} {2025}),\ \Eprint {https://arxiv.org/abs/2505.17583} {arXiv:2505.17583 [gr-qc]} \BibitemShut {NoStop}%
\bibitem [{\citenamefont {Awais}\ and\ \citenamefont {Azam}(2025)}]{awais2025anisotropiccompactstarvaidyatikekar}%
  \BibitemOpen
  \bibfield  {author} {\bibinfo {author} {\bibfnamefont {M.}~\bibnamefont {Awais}}\ and\ \bibinfo {author} {\bibfnamefont {M.}~\bibnamefont {Azam}},\ }\href {https://arxiv.org/abs/2503.20792} {\bibinfo {title} {Anisotropic compact star with vaidya-tikekar potential in $f(q)$ gravity}} (\bibinfo {year} {2025}),\ \Eprint {https://arxiv.org/abs/2503.20792} {arXiv:2503.20792 [gr-qc]} \BibitemShut {NoStop}%
\bibitem [{\citenamefont {Maurya}\ \emph {et~al.}(2023{\natexlab{b}})\citenamefont {Maurya}, \citenamefont {Errehymy}, \citenamefont {Jasim}, \citenamefont {Daoud}, \citenamefont {Al-Harbi},\ and\ \citenamefont {Abdel-Aty}}]{Maurya2023}%
  \BibitemOpen
  \bibfield  {author} {\bibinfo {author} {\bibfnamefont {S.~K.}\ \bibnamefont {Maurya}}, \bibinfo {author} {\bibfnamefont {A.}~\bibnamefont {Errehymy}}, \bibinfo {author} {\bibfnamefont {M.~K.}\ \bibnamefont {Jasim}}, \bibinfo {author} {\bibfnamefont {M.}~\bibnamefont {Daoud}}, \bibinfo {author} {\bibfnamefont {N.}~\bibnamefont {Al-Harbi}},\ and\ \bibinfo {author} {\bibfnamefont {A.-H.}\ \bibnamefont {Abdel-Aty}},\ }\href {https://doi.org/10.1140/epjc/s10052-023-11447-5} {\bibfield  {journal} {\bibinfo  {journal} {The European Physical Journal C}\ }\textbf {\bibinfo {volume} {83}},\ \bibinfo {pages} {317} (\bibinfo {year} {2023}{\natexlab{b}})}\BibitemShut {NoStop}%
\bibitem [{\citenamefont {Ditta}\ \emph {et~al.}(2023)\citenamefont {Ditta}, \citenamefont {Tiecheng}, \citenamefont {Errehymy}, \citenamefont {Mustafa},\ and\ \citenamefont {Maurya}}]{Ditta2023}%
  \BibitemOpen
  \bibfield  {author} {\bibinfo {author} {\bibfnamefont {A.}~\bibnamefont {Ditta}}, \bibinfo {author} {\bibfnamefont {X.}~\bibnamefont {Tiecheng}}, \bibinfo {author} {\bibfnamefont {A.}~\bibnamefont {Errehymy}}, \bibinfo {author} {\bibfnamefont {G.}~\bibnamefont {Mustafa}},\ and\ \bibinfo {author} {\bibfnamefont {S.~K.}\ \bibnamefont {Maurya}},\ }\href {https://doi.org/10.1140/epjc/s10052-023-11390-5} {\bibfield  {journal} {\bibinfo  {journal} {The European Physical Journal C}\ }\textbf {\bibinfo {volume} {83}},\ \bibinfo {pages} {254} (\bibinfo {year} {2023})}\BibitemShut {NoStop}%
\bibitem [{\citenamefont {Maurya}\ \emph {et~al.}(2022{\natexlab{b}})\citenamefont {Maurya}, \citenamefont {Mustafa}, \citenamefont {Govender},\ and\ \citenamefont {Newton~Singh}}]{Maurya_2022vsn}%
  \BibitemOpen
  \bibfield  {author} {\bibinfo {author} {\bibfnamefont {S.~K.}\ \bibnamefont {Maurya}}, \bibinfo {author} {\bibfnamefont {G.}~\bibnamefont {Mustafa}}, \bibinfo {author} {\bibfnamefont {M.}~\bibnamefont {Govender}},\ and\ \bibinfo {author} {\bibfnamefont {K.}~\bibnamefont {Newton~Singh}},\ }\href {https://doi.org/10.1088/1475-7516/2022/10/003} {\bibfield  {journal} {\bibinfo  {journal} {JCAP}\ }\textbf {\bibinfo {volume} {10}},\ \bibinfo {pages} {003}},\ \Eprint {https://arxiv.org/abs/2207.02021} {arXiv:2207.02021 [gr-qc]} \BibitemShut {NoStop}%
\bibitem [{\citenamefont {Calzá}\ and\ \citenamefont {Sebastiani}(2023)}]{Calz__2023}%
  \BibitemOpen
  \bibfield  {author} {\bibinfo {author} {\bibfnamefont {M.}~\bibnamefont {Calzá}}\ and\ \bibinfo {author} {\bibfnamefont {L.}~\bibnamefont {Sebastiani}},\ }\bibfield  {journal} {\bibinfo  {journal} {The European Physical Journal C}\ }\textbf {\bibinfo {volume} {83}},\ \href {https://doi.org/10.1140/epjc/s10052-023-11393-2} {10.1140/epjc/s10052-023-11393-2} (\bibinfo {year} {2023})\BibitemShut {NoStop}%
\bibitem [{\citenamefont {Bhar}\ \emph {et~al.}(2024)\citenamefont {Bhar}, \citenamefont {Shahzad}, \citenamefont {Mandal},\ and\ \citenamefont {Sahoo}}]{Bhar_2024}%
  \BibitemOpen
  \bibfield  {author} {\bibinfo {author} {\bibfnamefont {P.}~\bibnamefont {Bhar}}, \bibinfo {author} {\bibfnamefont {M.}~\bibnamefont {Shahzad}}, \bibinfo {author} {\bibfnamefont {S.}~\bibnamefont {Mandal}},\ and\ \bibinfo {author} {\bibfnamefont {P.}~\bibnamefont {Sahoo}},\ }\href {https://doi.org/10.1016/j.dark.2024.101686} {\bibfield  {journal} {\bibinfo  {journal} {Physics of the Dark Universe}\ }\textbf {\bibinfo {volume} {46}},\ \bibinfo {pages} {101686} (\bibinfo {year} {2024})}\BibitemShut {NoStop}%
\bibitem [{\citenamefont {Sokoliuk}\ \emph {et~al.}(2022)\citenamefont {Sokoliuk}, \citenamefont {Pradhan}, \citenamefont {Sahoo},\ and\ \citenamefont {Baransky}}]{Sokoliuk_2022}%
  \BibitemOpen
  \bibfield  {author} {\bibinfo {author} {\bibfnamefont {O.}~\bibnamefont {Sokoliuk}}, \bibinfo {author} {\bibfnamefont {S.}~\bibnamefont {Pradhan}}, \bibinfo {author} {\bibfnamefont {P.~K.}\ \bibnamefont {Sahoo}},\ and\ \bibinfo {author} {\bibfnamefont {A.}~\bibnamefont {Baransky}},\ }\bibfield  {journal} {\bibinfo  {journal} {The European Physical Journal Plus}\ }\textbf {\bibinfo {volume} {137}},\ \href {https://doi.org/10.1140/epjp/s13360-022-03273-7} {10.1140/epjp/s13360-022-03273-7} (\bibinfo {year} {2022})\BibitemShut {NoStop}%
\bibitem [{\citenamefont {Wang}\ \emph {et~al.}(2022)\citenamefont {Wang}, \citenamefont {Chen},\ and\ \citenamefont {Katsuragawa}}]{PhysRevD.105.024060}%
  \BibitemOpen
  \bibfield  {author} {\bibinfo {author} {\bibfnamefont {W.}~\bibnamefont {Wang}}, \bibinfo {author} {\bibfnamefont {H.}~\bibnamefont {Chen}},\ and\ \bibinfo {author} {\bibfnamefont {T.}~\bibnamefont {Katsuragawa}},\ }\href {https://doi.org/10.1103/PhysRevD.105.024060} {\bibfield  {journal} {\bibinfo  {journal} {Phys. Rev. D}\ }\textbf {\bibinfo {volume} {105}},\ \bibinfo {pages} {024060} (\bibinfo {year} {2022})}\BibitemShut {NoStop}%
\bibitem [{\citenamefont {Bhar}\ \emph {et~al.}(2023)\citenamefont {Bhar}, \citenamefont {Pradhan}, \citenamefont {Malik},\ and\ \citenamefont {Sahoo}}]{Bhar_2023}%
  \BibitemOpen
  \bibfield  {author} {\bibinfo {author} {\bibfnamefont {P.}~\bibnamefont {Bhar}}, \bibinfo {author} {\bibfnamefont {S.}~\bibnamefont {Pradhan}}, \bibinfo {author} {\bibfnamefont {A.}~\bibnamefont {Malik}},\ and\ \bibinfo {author} {\bibfnamefont {P.~K.}\ \bibnamefont {Sahoo}},\ }\bibfield  {journal} {\bibinfo  {journal} {The European Physical Journal C}\ }\textbf {\bibinfo {volume} {83}},\ \href {https://doi.org/10.1140/epjc/s10052-023-11745-y} {10.1140/epjc/s10052-023-11745-y} (\bibinfo {year} {2023})\BibitemShut {NoStop}%
\bibitem [{\citenamefont {Iqbal}\ \emph {et~al.}(2025)\citenamefont {Iqbal}, \citenamefont {Khan}, \citenamefont {Alshammari}, \citenamefont {Mohammed},\ and\ \citenamefont {Ilyas}}]{Iqbal:2025iln}%
  \BibitemOpen
  \bibfield  {author} {\bibinfo {author} {\bibfnamefont {N.}~\bibnamefont {Iqbal}}, \bibinfo {author} {\bibfnamefont {S.}~\bibnamefont {Khan}}, \bibinfo {author} {\bibfnamefont {M.}~\bibnamefont {Alshammari}}, \bibinfo {author} {\bibfnamefont {W.~W.}\ \bibnamefont {Mohammed}},\ and\ \bibinfo {author} {\bibfnamefont {M.}~\bibnamefont {Ilyas}},\ }\href {https://doi.org/10.1140/epjc/s10052-025-14102-3} {\bibfield  {journal} {\bibinfo  {journal} {Eur. Phys. J. C}\ }\textbf {\bibinfo {volume} {85}},\ \bibinfo {pages} {372} (\bibinfo {year} {2025})}\BibitemShut {NoStop}%
\bibitem [{\citenamefont {Paul}\ \emph {et~al.}(2025)\citenamefont {Paul}, \citenamefont {Kumar},\ and\ \citenamefont {Maurya}}]{paul2025studyphysicalpropertiescharacteristics}%
  \BibitemOpen
  \bibfield  {author} {\bibinfo {author} {\bibfnamefont {S.}~\bibnamefont {Paul}}, \bibinfo {author} {\bibfnamefont {J.}~\bibnamefont {Kumar}},\ and\ \bibinfo {author} {\bibfnamefont {S.~K.}\ \bibnamefont {Maurya}},\ }\href {https://arxiv.org/abs/2505.15853} {\bibinfo {title} {Study on physical properties and characteristics of an anisotropic compact star model using karmarkar condition in f(q) gravity}} (\bibinfo {year} {2025}),\ \Eprint {https://arxiv.org/abs/2505.15853} {arXiv:2505.15853 [gr-qc]} \BibitemShut {NoStop}%
\bibitem [{\citenamefont {Das}\ and\ \citenamefont {Paul}(2025)}]{das2025studypulsarexo1745248}%
  \BibitemOpen
  \bibfield  {author} {\bibinfo {author} {\bibfnamefont {B.}~\bibnamefont {Das}}\ and\ \bibinfo {author} {\bibfnamefont {B.~C.}\ \bibnamefont {Paul}},\ }\href {https://arxiv.org/abs/2509.11665} {\bibinfo {title} {A study of the pulsar exo 1745-248 in $f(q)$ gravity with pseudo-spheroidal geometry}} (\bibinfo {year} {2025}),\ \Eprint {https://arxiv.org/abs/2509.11665} {arXiv:2509.11665 [gr-qc]} \BibitemShut {NoStop}%
\bibitem [{\citenamefont {Sharif}\ and\ \citenamefont {Ibrar}(2025{\natexlab{a}})}]{sharif2025chargedanisotropicpulsarsax}%
  \BibitemOpen
  \bibfield  {author} {\bibinfo {author} {\bibfnamefont {M.}~\bibnamefont {Sharif}}\ and\ \bibinfo {author} {\bibfnamefont {I.}~\bibnamefont {Ibrar}},\ }\href {https://arxiv.org/abs/2505.00758} {\bibinfo {title} {Charged anisotropic pulsar sax j1748.9-2021 in non-riemannian geometry}} (\bibinfo {year} {2025}{\natexlab{a}}),\ \Eprint {https://arxiv.org/abs/2505.00758} {arXiv:2505.00758 [gr-qc]} \BibitemShut {NoStop}%
\bibitem [{\citenamefont {Sharif}\ and\ \citenamefont {Ibrar}(2025{\natexlab{b}})}]{sharif2025impactpulsarsaxj174892021}%
  \BibitemOpen
  \bibfield  {author} {\bibinfo {author} {\bibfnamefont {M.}~\bibnamefont {Sharif}}\ and\ \bibinfo {author} {\bibfnamefont {I.}~\bibnamefont {Ibrar}},\ }\href {https://arxiv.org/abs/2509.02641} {\bibinfo {title} {Impact of pulsar sax j1748.9-2021 observations on $f(\mathcal{Q}, \mathbb{T})$ gravity}} (\bibinfo {year} {2025}{\natexlab{b}}),\ \Eprint {https://arxiv.org/abs/2509.02641} {arXiv:2509.02641 [gr-qc]} \BibitemShut {NoStop}%
\bibitem [{\citenamefont {Narawade}\ \emph {et~al.}(2025)\citenamefont {Narawade}, \citenamefont {Lohakare},\ and\ \citenamefont {Mishra}}]{Narawade_2025}%
  \BibitemOpen
  \bibfield  {author} {\bibinfo {author} {\bibfnamefont {S.}~\bibnamefont {Narawade}}, \bibinfo {author} {\bibfnamefont {S.~V.}\ \bibnamefont {Lohakare}},\ and\ \bibinfo {author} {\bibfnamefont {B.}~\bibnamefont {Mishra}},\ }\href {https://doi.org/10.1016/j.aop.2024.169913} {\bibfield  {journal} {\bibinfo  {journal} {Annals of Physics}\ }\textbf {\bibinfo {volume} {474}},\ \bibinfo {pages} {169913} (\bibinfo {year} {2025})}\BibitemShut {NoStop}%
\bibitem [{\citenamefont {Abebe}\ \emph {et~al.}(2025)\citenamefont {Abebe}, \citenamefont {Apostolopoulos}, \citenamefont {Giacomini}, \citenamefont {Leon}, \citenamefont {Moncada},\ and\ \citenamefont {Paliathanasis}}]{abebe2025noncoincidencefq}%
  \BibitemOpen
  \bibfield  {author} {\bibinfo {author} {\bibfnamefont {A.}~\bibnamefont {Abebe}}, \bibinfo {author} {\bibfnamefont {P.~S.}\ \bibnamefont {Apostolopoulos}}, \bibinfo {author} {\bibfnamefont {A.}~\bibnamefont {Giacomini}}, \bibinfo {author} {\bibfnamefont {G.}~\bibnamefont {Leon}}, \bibinfo {author} {\bibfnamefont {F.}~\bibnamefont {Moncada}},\ and\ \bibinfo {author} {\bibfnamefont {A.}~\bibnamefont {Paliathanasis}},\ }\href {https://arxiv.org/abs/2510.00535} {\bibinfo {title} {Noncoincidence $f(q)$-cosmology with dark matter coupled to gravity}} (\bibinfo {year} {2025}),\ \Eprint {https://arxiv.org/abs/2510.00535} {arXiv:2510.00535 [gr-qc]} \BibitemShut {NoStop}%
\bibitem [{\citenamefont {De}\ and\ \citenamefont {Paliathanasis}(2025)}]{de2025exactcosmo}%
  \BibitemOpen
  \bibfield  {author} {\bibinfo {author} {\bibfnamefont {A.}~\bibnamefont {De}}\ and\ \bibinfo {author} {\bibfnamefont {A.}~\bibnamefont {Paliathanasis}},\ }\href {https://arxiv.org/abs/2509.14174} {\bibinfo {title} {Exact cosmological solutions in non-coincidence $f(q)$-theory}} (\bibinfo {year} {2025}),\ \Eprint {https://arxiv.org/abs/2509.14174} {arXiv:2509.14174 [gr-qc]} \BibitemShut {NoStop}%
\bibitem [{\citenamefont {Ayuso}\ \emph {et~al.}(2025)\citenamefont {Ayuso}, \citenamefont {Bouhmadi-López}, \citenamefont {Chen}, \citenamefont {Chew}, \citenamefont {Dialektopoulos},\ and\ \citenamefont {Ong}}]{ayuso2025insightsfqcosmologyrelevance}%
  \BibitemOpen
  \bibfield  {author} {\bibinfo {author} {\bibfnamefont {I.}~\bibnamefont {Ayuso}}, \bibinfo {author} {\bibfnamefont {M.}~\bibnamefont {Bouhmadi-López}}, \bibinfo {author} {\bibfnamefont {C.-Y.}\ \bibnamefont {Chen}}, \bibinfo {author} {\bibfnamefont {X.~Y.}\ \bibnamefont {Chew}}, \bibinfo {author} {\bibfnamefont {K.}~\bibnamefont {Dialektopoulos}},\ and\ \bibinfo {author} {\bibfnamefont {Y.~C.}\ \bibnamefont {Ong}},\ }\href {https://arxiv.org/abs/2506.03506} {\bibinfo {title} {Insights in $f(q)$ cosmology: the relevance of the connection}} (\bibinfo {year} {2025}),\ \Eprint {https://arxiv.org/abs/2506.03506} {arXiv:2506.03506 [gr-qc]} \BibitemShut {NoStop}%
\bibitem [{\citenamefont {Bahamonde}\ and\ \citenamefont {Järv}(2022)}]{Bahamonde_2022}%
  \BibitemOpen
  \bibfield  {author} {\bibinfo {author} {\bibfnamefont {S.}~\bibnamefont {Bahamonde}}\ and\ \bibinfo {author} {\bibfnamefont {L.}~\bibnamefont {Järv}},\ }\bibfield  {journal} {\bibinfo  {journal} {The European Physical Journal C}\ }\textbf {\bibinfo {volume} {82}},\ \href {https://doi.org/10.1140/epjc/s10052-022-10922-9} {10.1140/epjc/s10052-022-10922-9} (\bibinfo {year} {2022})\BibitemShut {NoStop}%
\bibitem [{\citenamefont {Dimakis}\ \emph {et~al.}(2025)\citenamefont {Dimakis}, \citenamefont {Giacomini}, \citenamefont {Paliathanasis},\ and\ \citenamefont {Panotopoulos}}]{dimakis2025relativisticstarsfqgravity}%
  \BibitemOpen
  \bibfield  {author} {\bibinfo {author} {\bibfnamefont {N.}~\bibnamefont {Dimakis}}, \bibinfo {author} {\bibfnamefont {A.}~\bibnamefont {Giacomini}}, \bibinfo {author} {\bibfnamefont {A.}~\bibnamefont {Paliathanasis}},\ and\ \bibinfo {author} {\bibfnamefont {G.}~\bibnamefont {Panotopoulos}},\ }\href {https://arxiv.org/abs/2503.14302} {\bibinfo {title} {Relativistic stars in $f(q)$-gravity}} (\bibinfo {year} {2025}),\ \Eprint {https://arxiv.org/abs/2503.14302} {arXiv:2503.14302 [gr-qc]} \BibitemShut {NoStop}%
\bibitem [{\citenamefont {Dimakis}\ \emph {et~al.}(2024)\citenamefont {Dimakis}, \citenamefont {Terzis}, \citenamefont {Paliathanasis},\ and\ \citenamefont {Christodoulakis}}]{dimakis2024staticsphericallysymmetricsolutions}%
  \BibitemOpen
  \bibfield  {author} {\bibinfo {author} {\bibfnamefont {N.}~\bibnamefont {Dimakis}}, \bibinfo {author} {\bibfnamefont {P.~A.}\ \bibnamefont {Terzis}}, \bibinfo {author} {\bibfnamefont {A.}~\bibnamefont {Paliathanasis}},\ and\ \bibinfo {author} {\bibfnamefont {T.}~\bibnamefont {Christodoulakis}},\ }\href {https://arxiv.org/abs/2410.04513} {\bibinfo {title} {Static, spherically symmetric solutions in $f(q)$-gravity and in nonmetricity scalar-tensor theory}} (\bibinfo {year} {2024}),\ \Eprint {https://arxiv.org/abs/2410.04513} {arXiv:2410.04513 [gr-qc]} \BibitemShut {NoStop}%
\bibitem [{\citenamefont {Bahamonde}\ \emph {et~al.}(2022)\citenamefont {Bahamonde}, \citenamefont {Gigante~Valcarcel}, \citenamefont {Järv},\ and\ \citenamefont {Lember}}]{Bahamonde_2022_2}%
  \BibitemOpen
  \bibfield  {author} {\bibinfo {author} {\bibfnamefont {S.}~\bibnamefont {Bahamonde}}, \bibinfo {author} {\bibfnamefont {J.}~\bibnamefont {Gigante~Valcarcel}}, \bibinfo {author} {\bibfnamefont {L.}~\bibnamefont {Järv}},\ and\ \bibinfo {author} {\bibfnamefont {J.}~\bibnamefont {Lember}},\ }\href {https://doi.org/10.1088/1475-7516/2022/08/082} {\bibfield  {journal} {\bibinfo  {journal} {Journal of Cosmology and Astroparticle Physics}\ }\textbf {\bibinfo {volume} {2022}}\bibinfo  {number} { (08)},\ \bibinfo {pages} {082}}\BibitemShut {NoStop}%
\bibitem [{\citenamefont {Järv}\ \emph {et~al.}(2018)\citenamefont {Järv}, \citenamefont {Rünkla}, \citenamefont {Saal},\ and\ \citenamefont {Vilson}}]{Jrv2018}%
  \BibitemOpen
\bibfield  {number} {  }\bibfield  {author} {\bibinfo {author} {\bibfnamefont {L.}~\bibnamefont {Järv}}, \bibinfo {author} {\bibfnamefont {M.}~\bibnamefont {Rünkla}}, \bibinfo {author} {\bibfnamefont {M.}~\bibnamefont {Saal}},\ and\ \bibinfo {author} {\bibfnamefont {O.}~\bibnamefont {Vilson}},\ }\bibfield  {journal} {\bibinfo  {journal} {Physical Review D}\ }\textbf {\bibinfo {volume} {97}},\ \href {https://doi.org/10.1103/physrevd.97.124025} {10.1103/physrevd.97.124025} (\bibinfo {year} {2018})\BibitemShut {NoStop}%
\bibitem [{\citenamefont {Beltr{\'a}n~Jim{\'e}nez}\ \emph {et~al.}(2020{\natexlab{b}})\citenamefont {Beltr{\'a}n~Jim{\'e}nez}, \citenamefont {Heisenberg},\ and\ \citenamefont {Koivisto}}]{BeltranJimenez:2020sih}%
  \BibitemOpen
  \bibfield  {author} {\bibinfo {author} {\bibfnamefont {J.}~\bibnamefont {Beltr{\'a}n~Jim{\'e}nez}}, \bibinfo {author} {\bibfnamefont {L.}~\bibnamefont {Heisenberg}},\ and\ \bibinfo {author} {\bibfnamefont {T.}~\bibnamefont {Koivisto}},\ }\href {https://doi.org/10.1088/1361-6382/aba31b} {\bibfield  {journal} {\bibinfo  {journal} {Class. Quant. Grav.}\ }\textbf {\bibinfo {volume} {37}},\ \bibinfo {pages} {195013} (\bibinfo {year} {2020}{\natexlab{b}})},\ \Eprint {https://arxiv.org/abs/2004.04606} {arXiv:2004.04606 [hep-th]} \BibitemShut {NoStop}%
\bibitem [{\citenamefont {Jim{\'{e}}nez}\ \emph {et~al.}(2020)\citenamefont {Jim{\'{e}}nez}, \citenamefont {Heisenberg}, \citenamefont {Koivisto},\ and\ \citenamefont {Pekar}}]{Jim_nez_2020}%
  \BibitemOpen
  \bibfield  {author} {\bibinfo {author} {\bibfnamefont {J.~B.}\ \bibnamefont {Jim{\'{e}}nez}}, \bibinfo {author} {\bibfnamefont {L.}~\bibnamefont {Heisenberg}}, \bibinfo {author} {\bibfnamefont {T.}~\bibnamefont {Koivisto}},\ and\ \bibinfo {author} {\bibfnamefont {S.}~\bibnamefont {Pekar}},\ }\bibfield  {journal} {\bibinfo  {journal} {Physical Review D}\ }\textbf {\bibinfo {volume} {101}},\ \href {https://doi.org/10.1103/physrevd.101.103507} {10.1103/physrevd.101.103507} (\bibinfo {year} {2020})\BibitemShut {NoStop}%
\bibitem [{\citenamefont {Zhao}(2022)}]{Zhao_2022}%
  \BibitemOpen
  \bibfield  {author} {\bibinfo {author} {\bibfnamefont {D.}~\bibnamefont {Zhao}},\ }\bibfield  {journal} {\bibinfo  {journal} {The European Physical Journal C}\ }\textbf {\bibinfo {volume} {82}},\ \href {https://doi.org/10.1140/epjc/s10052-022-10266-4} {10.1140/epjc/s10052-022-10266-4} (\bibinfo {year} {2022})\BibitemShut {NoStop}%
\bibitem [{\citenamefont {Misner}\ \emph {et~al.}(2017)\citenamefont {Misner}, \citenamefont {Thorne}, \citenamefont {Wheeler},\ and\ \citenamefont {Kaiser}}]{misner2017gravitation}%
  \BibitemOpen
  \bibfield  {author} {\bibinfo {author} {\bibfnamefont {C.}~\bibnamefont {Misner}}, \bibinfo {author} {\bibfnamefont {K.}~\bibnamefont {Thorne}}, \bibinfo {author} {\bibfnamefont {J.}~\bibnamefont {Wheeler}},\ and\ \bibinfo {author} {\bibfnamefont {D.}~\bibnamefont {Kaiser}},\ }\href {https://books.google.de/books?id=zAAuDwAAQBAJ} {\emph {\bibinfo {title} {Gravitation}}}\ (\bibinfo  {publisher} {Princeton University Press},\ \bibinfo {year} {2017})\BibitemShut {NoStop}%
\bibitem [{\citenamefont {Schutz}(2009)}]{schutz2009first}%
  \BibitemOpen
  \bibfield  {author} {\bibinfo {author} {\bibfnamefont {B.}~\bibnamefont {Schutz}},\ }\href {https://books.google.de/books?id=V1CGLi58W7wC} {\emph {\bibinfo {title} {A First Course in General Relativity}}}\ (\bibinfo  {publisher} {Cambridge University Press},\ \bibinfo {year} {2009})\BibitemShut {NoStop}%
\bibitem [{\citenamefont {O'Boyle}\ \emph {et~al.}(2020)\citenamefont {O'Boyle}, \citenamefont {Markakis}, \citenamefont {Stergioulas},\ and\ \citenamefont {Read}}]{O_Boyle_2020}%
  \BibitemOpen
  \bibfield  {author} {\bibinfo {author} {\bibfnamefont {M.~F.}\ \bibnamefont {O'Boyle}}, \bibinfo {author} {\bibfnamefont {C.}~\bibnamefont {Markakis}}, \bibinfo {author} {\bibfnamefont {N.}~\bibnamefont {Stergioulas}},\ and\ \bibinfo {author} {\bibfnamefont {J.~S.}\ \bibnamefont {Read}},\ }\bibfield  {journal} {\bibinfo  {journal} {Physical Review D}\ }\textbf {\bibinfo {volume} {102}},\ \href {https://doi.org/10.1103/physrevd.102.083027} {10.1103/physrevd.102.083027} (\bibinfo {year} {2020})\BibitemShut {NoStop}%
\bibitem [{\citenamefont {Read}\ \emph {et~al.}(2009)\citenamefont {Read}, \citenamefont {Lackey}, \citenamefont {Owen},\ and\ \citenamefont {Friedman}}]{Read_2009}%
  \BibitemOpen
  \bibfield  {author} {\bibinfo {author} {\bibfnamefont {J.~S.}\ \bibnamefont {Read}}, \bibinfo {author} {\bibfnamefont {B.~D.}\ \bibnamefont {Lackey}}, \bibinfo {author} {\bibfnamefont {B.~J.}\ \bibnamefont {Owen}},\ and\ \bibinfo {author} {\bibfnamefont {J.~L.}\ \bibnamefont {Friedman}},\ }\bibfield  {journal} {\bibinfo  {journal} {Physical Review D}\ }\textbf {\bibinfo {volume} {79}},\ \href {https://doi.org/10.1103/physrevd.79.124032} {10.1103/physrevd.79.124032} (\bibinfo {year} {2009})\BibitemShut {NoStop}%
\end{thebibliography}%

\end{document}